\documentclass[prd,nofootinbib,12pt,amsart]{revtex4}
\usepackage[normalem]{ulem}
\usepackage{soul}
\usepackage[utf8]{inputenc}
\usepackage{lineno,hyperref}
\usepackage{amsfonts,amssymb,amsmath}
\usepackage{epsfig}
\usepackage{mathrsfs}
\usepackage{amssymb}
\usepackage{graphicx}
\usepackage{amsmath}

\usepackage{xcolor}
\usepackage{color}
\usepackage[normalem]{ulem}
\usepackage{soul}
\usepackage{appendix}
\usepackage{amsmath}

\usepackage{xcolor} 
\begin{document}

\title{Circuit and Krylov complexity of primordial perturbations of modified gravity in inflation}
\author{Tao Li}
\email{2022700450@stu.jsu.edu.cn}
\author{Hai-Bing Fu}
 \affiliation{Department of Physics, Guizhou Minzu University, Guiyang 550025, China}

\begin{abstract}
Taking canonical scalar field inflation as the reference case, we investigate the circuit and Krylov complexities of primordial curvature perturbations in an \(f(\phi,R)\) modified gravity inflation. We derive the evolution equations for the squeezed strength \(r_k\) and squeezed angle \(\varphi_k\), and calculate the circuit complexity, Krylov complexity, Krylov entropy, Lanczos coefficients, and effective dissipative contribution. Compared with the canonical model, the squeezed strength \(r_k\) grows more slowly in the modified gravity inflation and continues to exhibit oscillations after horizon exit. Consequently, the spreading in Krylov space is suppressed and modulated through the relation \(K=\sinh^2 r_k\). Modified gravity also changes the squeezed angle, thereby modifying the phase-space correlations and enhancing the oscillatory behavior of the circuit complexity. In the presence of modified gravity effects, information loss is suppressed before horizon exit, whereas it barely affects information loss after horizon exit. For the parameters considered, the Lyapunov exponent and Lanczos coefficients indicate that modified gravity has little effect on the effective degree of quantum chaoticity. These results show that \(f(\phi,R)\) modified gravity primarily reorganizes the amplitude, phase, complexity, and information loss dynamics of primordial quantum fluctuations.
\end{abstract}

\maketitle

\section{introduction}
\label{introduction}
Understanding the observational consequences of inflationary quantum fluctuations is a central theme in modern cosmology, particularly due to their role in seeding large-scale structures and primordial black holes. Nevertheless, the microscopic origin of these fluctuations remains elusive. Insights from the holographic principle suggest that spacetime emerges fundamentally from quantum entanglement \cite{VanRaamsdonk:2010pw}, as exemplified by the Einstein-Rosen bridge correspondence \cite{Maldacena:2013xja}. Under this framework, a well-known puzzle arises: the boundary conformal field theory (CFT) thermalizes rapidly, whereas the bulk wormhole geometry evolves over a much longer timescale \cite{Stanford:2014jda}. This discrepancy directly motivated the introduction of quantum complexity to characterize the long-time dynamics of the holographic bulk \cite{Hartman:2013qma,Liu:2013iza}.

Primordial quantum fluctuations can be effectively probed using circuit complexity and Krylov complexity. Rooted in quantum computation \cite{Bhattacharyya:2020rpy,Liu:2021nzx,Li:2021kfq,Li:2023ekd,Li:2021sro}, the core formalism of circuit complexity has been widely adapted to cosmological scenarios.We take the state of a selected pair of perturbation modes as the target and evaluate its preparation cost relative to a specified reference state. In holography, quantum complexity is conjectured to be dual to the time evolution of wormholes \cite{Hartman:2013qma}. This characterization was formalized by Susskind et al.\ through the holographic complexity conjectures \cite{Susskind:2014moa}—specifically the complexity=volume (CV) and complexity=action (CA) proposals—which map complexity to either the maximal spatial bulk volume or the gravitational action within the Wheeler-DeWitt patch \cite{Susskind:2014rva}. In contrast, Krylov complexity diagnoses the operator growth by the Hamiltonian of the system \cite{Adhikari:2022oxr}, offering unique insights into the dynamical features and physical mechanisms that drive early-universe quantum fluctuations \cite{Li:2024kfm,Liu:2025caj,Zhai:2024odw}. Although saturation has only been established for small precursors in generic systems, Ref. \cite{Craps:2025kub} show exact Krylov complexity–Nielsen complexity squared matching over a finite precursor range in the Sachdev–Ye–Kitaev (SYK) model.

First, circuit complexity \cite{Aaronson:2016vto} is defined as the minimum number of unitary operators required to transition from a reference state to a target state, or equivalently, the minimal number of operations needed to complete a specific task. There are two primary methodologies for analyzing circuit complexity: the geometric methods developed by Nielsen et al. within the quantum gate complexity space \cite{Nielsen:2005mkt,Nielsen:2006cea,Dowling:2006tnk}, and the approach utilizing the Fubini-Study distance \cite{Chapman:2017rqy,Zhai:2024tkz}. Based on these frameworks, complexity can be explicitly derived from either the wave function \cite{Jefferson:2017sdb,Bhattacharyya:2018bbv,Jiang:2018nzg} or the covariance matrix \cite{Jiang:2018nzg,Alves:2018qfv,Camargo:2018eof,Ali:2018aon,Khan:2018rzm}. Furthermore, circuit complexity has been leveraged to elucidate the physics of Hawking radiation \cite{Zhao:2019nxk,Brown:2019rox} and to probe the emergent nature of spacetime \cite{Chandra:2021kdv}.

The second approach is based on the formalism of Krylov complexity. Krylov complexity was introduced as a diagnostic of operator growth in a Krylov basis, where the basis is generated recursively by the Liouvillian or Hamiltonian through the Lanczos algorithm. The geometric formulation of Krylov complexity and its relation to operator growth were further developed in Refs. \cite{Balasubramanian:2022tpr,Caputa:2021sib}. In particular, these works relate Krylov evolution to geometric motion on generalized coherent-state manifolds and to the spreading of quantum states over an appropriately chosen basis. Once a quantum state or operator is specified, the assciated Krylov space can be systematically constructed via the Lanczos algorithm \cite{Jafarizadeh:2006woc}, allowing for the explicit evaluation of the Krylov complexity, Lanczos coefficients, and Krylov entropy. Methodologies for computing the Krylov complexity of various orthogonal polynomials have been detailed in Ref. \cite{Muck:2022xfc}. Krylov complexity has been extensively applied to the SYK model \cite{Parker:2018yvk,Rabinovici:2020ryf,Jian:2020qpp,He:2022ryk, Basu:2025ubf}, and can be extended to a thermalized counterpart \cite{Kar:2021nbm}. Within this framework, quantum chaoticity itself can be interpreted as a non-local delocalization property in Krylov space \cite{Dymarsky:2019elm}. Recent advancements span a wide range of topics \cite{Erdmenger:2023wjg,Patramanis:2023cwz,Fan:2023ohh,Hashimoto:2023swv,Vasli:2023syq,Domingo:2023kjr,Gill:2023umm,Bhattacharjee:2023uwx,Adhikari:2022whf}. In particular, Ref. \cite{Camargo:2023eev, Bhattacharjee:2022vlt} investigated Krylov operator complexity in simple quantumsystems, while Ref. \cite{Huh:2023jxt} verified the universal characteristics of spread complexity in saddle-dominated scrambling scenarios. Furthermore, we have previously investigated non-trivial sound speed models and cosmological complexity within the context of k-essence \cite{Liu:2021nzx,Li:2021kfq}. 

The inflationary paradigm \cite{Linde:2007fr, Gorbunov:2011zzc, Lyth:1998xn, Linde:1983gd} is a well-established theoretical framework describing a period of exponential, accelerated expansion in the very early universe. As one of the most successful theories in modern cosmology, it consistently yields a nearly scale-invariant primordial power spectrum. Historically, the original approach to inflation was formulated using a single canonical scalar field undergoing slow-roll dynamics along its potential \cite{Linde:1993cn, Sasaki:1995aw, Turok:2002yq,Baumann:2014nda}. This paradigm successfully resolves three long-standing cosmological puzzles: 1) the horizon problem \cite{Guth:1980zm}, 2) the flatness problem \cite{Linde:1981mu}, and 3) the magnetic monopole problem \cite{Albrecht:1982wi}. Furthermore, its three core contributions include: 1) explaining the microscopic origin of temperature anisotropies in the cosmic microwave background (CMB) \cite{Mukhanov:1981xt,Starobinsky:1982ee}; 2) predicting a background of primordial gravitational waves \cite{Starobinsky:1979ty,Grishchuk:1974ny}; and 3) providing the primordial 'seeds' required for cosmic large-scale structure formation \cite{Guth:1982ec,Hawking:1982cz}. During the inflationary era, quantum fluctuations of the inflation field—which can be fundamentally understood as a scalar field—play a key role in seeding these subsequent structures. These microscopic quantum fluctuations were rapidly stretched to macroscopic physical scales by the cosmic expansion and subsequently transitioned into classical density perturbations through quantum decoherence. Driven by gravitational instability, these minute density inhomogeneities gradually grew over time, eventually evolving into the large-scale structure of the universe observed today.

As a baseline, we study quantum complexity within the inflationary paradigm \cite{Linde:2007fr,Gorbunov:2011zzc,Lyth:1998xn,Linde:1983gd,Linde:1993cn,Sasaki:1995aw,Turok:2002yq,Baumann:2014nda}, which is the simplest and most representative realization of this theoretical framework. In this model, the accelerated expansion of the Universe is driven primarily by the inflationary field \(\phi\). Quantum fluctuations generated on microscopic scales are subsequently stretched to cosmological scales by inflation, eventually becoming the seeds of large-scale structure formation. To investigate the effects of gravity on quantum complexity during inflation, we first consider \(F(R)\) gravity. Our analysis shows that, after an appropriate field transformation and reformulation, \(F(R)\) gravity can be reduced to an equivalent canonical scalar field inflation model. We therefore adopt it as a reference baseline in our study. Further analysis indicates that gravitational effects have a nontrivial impact on quantum complexity only when they are coupled to the scalar field. Motivated by this observation, we subsequently consider the modified gravity \(f(\phi,R)\) in order to systematically investigate the effects of gravity–scalar field coupling on quantum complexity. By systematically comparing the generalized model with the canonical one, we investigate how modified gravitational dynamics affect the statistical properties and evolution of primordial quantum fluctuations. Within this unified framework, we study circuit complexity and Krylov complexity during the inflationary epoch in order to characterize how the scalar–curvature coupling modifies the quantum dynamics of primordial perturbations

This paper is organized as follows. In Sec.~\ref{Krylov complexity and circuit complexity}, we will provide a brief overview of the theoretical foundations of both circuit complexity and Krylov complexity. In Sec.~\ref{Squeezed parameter in the inﬂationary paradigm}, we will derive the governing differential equations for the squeezed parameters $(r_k, \varphi_k)$ within the frameworks of canonical scalar field inflation and modified gravity $f(\phi,R)$ inflation. Sec.~\ref{evolution of Circuit complexity and Krylov complexity} will present the numerical evolution of circuit and Krylov complexity, along with their associated physical quantities, including the Lyapunov exponent, Krylov entropy, Lanczos coefficients $b_n$, and the dissipation coefficient $c_n$ for open quantum systems. Finally, concluding remarks and future outlooks will be provided in Sec.~\ref{conclusion}.

\section{Krylov complexity and circuit complexity}
\label{Krylov complexity and circuit complexity}

In this section, we provide a concise overview of the theoretical frameworks and mathematical formalisms underpinning both circuit complexity and Krylov complexity.

\subsection{Krylov complexity and Lanczos algorithm}
\label{Krylov complexity}

In this section,we distinguish operator Krylov complexity from state spreading and specify which construction is used in the calculations below. In the Heisenberg picture, the initial observable $\mathcal{O}$ evolves over time to
\begin{equation}\label{wave function of O}
    \mathcal{O} (t)=e^{it\mathcal{L}}\mathcal{O}=e^{iHt}\mathcal{O}e^{-iHt}=\sum_{n=0}^{\infty } \frac{(it)^{n}}{n!} \mathcal{L}^{n}\mathcal{O} (0)=\sum_{n=0}^{\infty } \frac{(it)^{n}}{n!} \tilde{\mathcal{O} }_{n} ,
\end{equation}
with
\begin{equation}
    \mathcal{L}=[H,.],~~ \mathcal{L}^{n}\mathcal{O} =\tilde{\mathcal{O} }_{n} =[H,\tilde{\mathcal{O}}_{n-1}], 
\end{equation}
where $\mathcal{L}$ is the Liouvillian superoperator and define $\mathcal{L}_{X}$ as $\mathcal{L}_{X}Y=[X,Y]$ . Clearly, $\mathcal{O}(t)$ describes a trajectory within the vector space of all Hermitian operators. In general, its complexity increases gradually. The time evolution of a dynamical operator $\mathcal{O}(t)$ is governed by the Heisenberg equation of motion $\partial _{t}\mathcal{O} (t)=i[H,\mathcal{O} (t)]$, the $H$ represents the Hamiltonian. Next, we can implement the Lanczos algorithm to construct an orthogonal Krylov basis, where the operators are \cite{Muck:2022xfc}
\begin{equation}
\begin{split}
|\mathcal{O}_{0})= |{\mathcal{O} }),~ |\mathcal{O}_{1})=b_1^{-1}|A_1) ~ ... ~ |\mathcal{O}_{n})=b_1^{-1}|A_n),
\end{split}
\label{krylov basis}
\end{equation}
with
\begin{equation}
\begin{split}
|A_{1})=\mathcal{L}|\mathcal{O}_{0}),~|A_{n})=\mathcal{L}|\mathcal{O}_{n-1})-b_{n-1} |\mathcal{O}_{n-2}).
\end{split}
\label{An}
\end{equation}
Where $b_{1}=\sqrt{(\tilde{\mathcal{O}}_{0}\mathcal{L}|\mathcal{L}\tilde{\mathcal{O}}_{0})  }$ describes the normalized vector. And $b_{n}=\sqrt{(A_{n}|A_{n})}$ is referred to as the Lanczos coefficients. Subsequently, we can expand Eq. \eqref{wave function of O} as $\sum_{n=0}^{\infty } (i)^{n} \Psi_{n}(t)|\mathcal{O}_{n})$, where $\Psi_n$ is the wave function and satisfies the normalization condition $\sum_{n}\left | \Psi_{n} \right |^{2}=1$. Substituting the wave function into the Schrödinger equation yields
\begin{equation}
\begin{split}
\partial _{t}\Psi_{n}(t)=b_{n}\Psi_{n-1}-b_{n-1}\Psi_{n+1}, ~\Psi_{n}(0)=\delta_{n0}. 
\end{split}
\label{eom of phi}
\end{equation}
Subsequently, the Krylov complexity can be defined as follows based on $\phi_n$
\begin{equation}
 \begin{split}
 K=\sum_{n=1}^{k_n}n\left | \Psi_{n} \right | ^{2},
 \end{split}\label{krylov complexity}
 \end{equation}
where $k_n$ denotes the dimension of the Krylov basis $\mathcal{O}_n$. Furthermore, in the subsequent discussion, the specific expression for Krylov complexity can be derived from this definition \ref{krylov complexity}.

The geometric formulation of Krylov complexity and its relation to operator growth have been developed in Refs. \cite{Balasubramanian:2022tpr,Caputa:2021sib}. In particular, these works establish connections between Krylov evolution, generalized coherent-state manifolds, geometric motion, and the spreading of quantum states. The present work applies this established framework to the two-mode squeezed state of primordial perturbations in canonical scalar field inflation and \(f(\phi,R)\) modified gravity.

\subsection{Circuit complexity}
\label{Circuit complexity}
Nielsen et al. employed a geometric approach to study the evolution of quantum complexity \cite{Nielsen:2005mkt,Nielsen:2006cea,Dowling:2006tnk}. The state $|\Psi^{R}\rangle$ at $\tau = 0$ is defined as the reference state, whereas the state $|\Psi^{T}\rangle$ at $\tau = 1$ is identified as the target state. The relation between them is given by
\begin{equation}\label{target state}
\left|\Psi^{T}\right\rangle_{\tau=1}=U(\tau=1)\left|\psi^{R}\right\rangle_{\tau=0},
\end{equation}
where $\tau$ is used to parameterize the Hilbert space in quantum mechanics. $U(\tau)$ is a unitary operator, which in quantum field theory can be constructed as the path-ordered exponential of the Hamiltonian operator
\begin{equation}\label{U tau and H}
    U(\tau)=\overleftarrow{\mathcal{P}} \exp \left(-i \int_{0}^{\tau} d s H(s)\right),
\end{equation}
where $\overleftarrow{\mathcal{P}}$ denotes path ordering from right to left. In this framework, the Hamiltonian is expanded in a basis of Hermitian operators $M_I$, which generate elementary quantum gates, as
$H(s) = Y^{I}(s)\, M_I$.
Here, $Y^{I}(s)$ acts as a control function specifying which gate is activated during the evolution. The corresponding unitary operator satisfies the Schrödinger equation $\frac{d U}{d s}=-i Y(s)^{I} M_{I} U(s)$. In order to define circuit complexity, functions related to circuit complexity should be introduced
\begin{equation}\label{minimizing the cost function}
    C(U)=\int_{0}^{1} \mathcal{F}(U, \dot{U}) d \tau.
\end{equation}
The circuit complexity is obtained by minimizing the cost function in Eq.~\ref{minimizing the cost function}. This procedure yields the shortest geodesic connecting the reference state and the target state. We first consider the quadratic cost function $\mathcal{F}(U, Y)=\sqrt{\sum_{I}(Y^{I})^{2}}$. The target wave function is a two-mode squeezed state, which in momentum space can be written as
\begin{equation}\label{target wave function}
    \begin{aligned}
\Psi_{\mathrm{sq}}\left(q_{\vec{k}}, q_{-\vec{k}}\right) & =\sum_{n=0}^{\infty}(-1)^{n} \frac{\tanh ^{n} r_{k}}{\cosh ^{n} r_{k}}\left\langle q_{\vec{k}} ; q_{-\vec{k}} \mid n ; n\right\rangle_{\vec{k},-\vec{k}} \\
& =\frac{\exp \left[A\left(r_{k}, \varphi_{k}\right) \cdot\left(q_{k}^{2}+q_{-k}^{2}\right)-B\left(r_{k}, \varphi_{k}\right) q_{\vec{k}} q_{-\vec{k}}\right]}{\cosh r_{k} \sqrt{\pi} \sqrt{1-e^{-4 i \varphi_{k}}} \tanh ^{2} r_{k}},
\end{aligned}
\end{equation}
where $A(r_k,\varphi_k)$ and $B(r_k,\varphi_k)$ are defined as follows
\begin{equation}
    A\left(r_{k}, \varphi_{k}\right)=\frac{k}{2}\left(\frac{e^{-4 i \varphi_{k}} \tanh ^{2} r_{k}+1}{e^{-4 i \varphi_{k}} \tanh ^{2} r_{k}-1}\right),~
B\left(r_{k},\varphi_{k}\right)=\frac{k}{2}\left(\frac{e^{-2 i \varphi_{k}} \tanh ^{2} r_{k}}{e^{-4 i \varphi_{k}} \tanh ^{2} r_{k}-1}\right).
\end{equation}
After performing an appropriate rotation of the pair $(\vec{q}_k, \vec{q}_{-k})$ in momentum space, Eq.~\ref{target wave function} can be cast into diagonal form. Consequently, the squeezed vacuum state serves as the reference state and can be expressed using the standard procedure as follows
\begin{equation}
\begin{split}
\Psi_{00}\left(q_{\vec{k}}, q_{-\vec{k}}\right) =\left\langle q_{\vec{k}} ; q_{-\vec{k}} \mid 0 ; 0\right\rangle_{\vec{k},-\vec{k}} 
=\frac{\exp \left[-\frac{1}{2} \tilde{M}^{a b} q_{a} q_{b}\right]}{\pi^{1 / 2}}    
\end{split},
\end{equation}
in which
\begin{equation}
  \tilde{M} = \begin{pmatrix}
 -2A\left(r_{k},\varphi_{k}\right)+B\left(r_{k}, \varphi_{k}\right) & 0\\
  0&-2A\left(r_{k}, \varphi_{k}\right)-B\left(r_{k}, \varphi_{k}\right)
\end{pmatrix}.
\end{equation}
According to Eq. \ref{target state}, $\Psi_{00}\left(q_{\vec{k}}, q_{-\vec{k}}\right)$ and $U(\tau)$ can be related to their corresponding target states via Eq. \ref{U tau and H}.
\begin{equation}
    \Psi_{\tau}\left(q_{\vec{k}}, q_{-\vec{k}}\right)=\tilde{U}(\tau) \Psi_{00}\left(q_{\vec{k}}, q_{-\vec{k}}\right) \tilde{U}^{\dagger}(\tau),
\end{equation}
\begin{equation}\label{the boundary condition}
    \Psi_{\tau=0}\left(q_{\vec{k}}, q_{-\vec{k}}\right)=\Psi_{00}\left(q_{\vec{k}}, q_{-\vec{k}}\right),~
\Psi_{\tau=1}\left(q_{\vec{k}}, q_{-\vec{k}}\right)=\Psi_{\mathrm{sq}}\left(q_{\vec{k}}, q_{-\vec{k}}\right).
\end{equation}
In particular, $U(\tau)$ is a unitary matrix in $GL(2,C)$, which describes the geodesic between the reference state and the target state in the parameter space. According to Ref. \cite{Jefferson:2017sdb}, $U(\tau)$ will take the following form
\begin{equation}
    \tilde{U}(\tau)=\exp \left[\sum_{k=1}^{2} Y^{k}(\tau) M_{k}^{\mathrm{diag}}\right],
\end{equation}
where $M_{k}^{\text {diag }}$ denotes two generators of $ G L(2, C)$  that defines as
\begin{equation}
    M_{1}^{\mathrm{diag}}=\left(\begin{array}{ll}
1 & 0 \\
0 & 0
\end{array}\right), M_{2}^{\mathrm{diag}}=\left(\begin{array}{ll}
0 & 0 \\
0 & 1
\end{array}\right) .
\end{equation}
The off-diagonal elements are zero. Since $U(\tau)$ can parameterize the geodesics in the group manifold, as it generates the non-trivial curvature of the group manifold. According to Ref. \cite{Ali:2018fcz}, $Y_I(\tau)$ is constructed as $Y_{I}(\tau)=Y_{I}(\tau=1) \cdot \tau+Y_{I}(\tau=0)$. From the boundary condition \ref{the boundary condition}, we obtain
\begin{equation}
    \begin{array}{c}
\left.\operatorname{Im}\left(\mathrm{Y}^{1,2}\right)\right|_{\tau=0}=\left.\operatorname{Re}\left(\mathrm{Y}^{\mathrm{I}}\right)\right|_{\tau=0}=0, ~
\left.\operatorname{Im}\left(\mathrm{Y}^{1,2}\right)\right|_{\tau=1}=\frac{1}{2} \ln \frac{\left|\Omega_{\tilde{\mathrm{k}},-\tilde{\mathrm{k}}}\right|}{\omega_{\tilde{\mathrm{k}},-\tilde{\mathrm{k}}}}, \\
\left.\operatorname{Re}\left(\mathrm{Y}^{1,2}\right)\right|_{\tau=1}=\frac{1}{2} \arctan \frac{\operatorname{Im}\left(\Omega_{\tilde{\mathrm{k}},-\tilde{\mathrm{k}}}\right)}{\operatorname{Re}\left(\omega_{\tilde{\mathrm{k}},-\tilde{\mathrm{k}}}\right)} .
\end{array}
\end{equation}
Given these conditions, we can express circuit complexity as a geodesic in the parameter manifold shown below
\begin{equation}\label{circuit complexity as a geodesic in the parameter manifold }
    C(\tilde{U})=\int_{0}^{1} d \tau \sqrt{G_{I J} \dot{Y}^{I}(\tau) \dot{Y}^{I}(\tau)^{*}},
\end{equation}
where \(G_{IJ}\) is the induced metric of the group manifold rather than that of spacetime. Ref. \cite{Jefferson:2017sdb} shows that \(G_{IJ}\) can possess any structure corresponding to the various structures of the group manifold. As mentioned earlier, the induced metric of our group structure is flat. Therefore, we can substitute $Y_{I}(\tau)=Y_{I}(\tau=1) \cdot \tau+Y_{I}(\tau=0)$ into Eq. \ref{circuit complexity as a geodesic in the parameter manifold } to obtain the circuit complexity
\begin{equation}\label{last circuit complexity}
    \begin{aligned}
C(k)& =\frac{1}{\sqrt{2}}\Bigg[\log\bigg(\frac{\cosh{}^{-4}{(r_k)}+\sin^2{(2\varphi_k)\tanh^2{(r_k)}}}{1-2\cos{(2\varphi_k)\tanh{r_k}+\tanh^2{r_k}}}\bigg)^{2}\\&+\bigg(\arctan{\big(2\sin{(2\varphi_k)\sinh{r_k\cosh{r_k}}}\big)} \bigg)^{2}\Bigg]^{\frac{1}{2}}
\end{aligned}
\end{equation}
\section{Squeezed parameter in two inflationary models}
\label{Squeezed parameter in the inﬂationary paradigm}

The inflationary paradigm is one of the most successful frameworks for describing the early universe, as it naturally generates a nearly scale-invariant power spectrum. In the cosmology literature, the inflationary epoch is commonly modeled by a single scalar field undergoing slow-roll evolution along its potential, which constitutes the original formulation of inflation \cite{Linde:2007fr,Gorbunov:2011zzc,Lyth:1998xn,Linde:1983gd,Linde:1993cn,Sasaki:1995aw,Turok:2002yq,Baumann:2014nda}. 

In this section, we provide a brief overview of canonical scalar field inflation and $f(\phi,R)$ inflation, without entering into excessive technical detail.Our focus is on the evolution of the canonical perturbation modes and the resulting Gaussian-state diagnostics. We begin with the canonical scalar field description and then proceed to the $f(\phi,R)$ framework.We restrict thecalculation tothe model defined below and to its specified scalar perturbation sector. A detailed discussion of cosmological perturbation theory can be found in Ref.~\cite{Mukhanov:1990me}, and we shall not repeat it here.

\subsection{Canonical inflation }
\label{Canonical scalar field inﬂation}

In this section, we consider the Friedman–Lemaitre-Robertson–Walker (FLRW) metric as follows,
\begin{equation}
    ds^2=a^2(\tau)(-d\tau^2+d\vec x^2),
    \label{frwl metric}
\end{equation}
where $a(\tau)$ is the scale factor in terms of conformal time $\tau$. To define the Mukhanov-Sasaki variable, we also need the perturbations of the FLRW metric \eqref{frwl metric}, which are given by
\begin{equation}
    ds^2= a^2(\tau)\left[-(1+\Psi(x_\mu))d\tau^2+(1-\Psi(x_\mu))d\vec{x}^2\right]
\end{equation}
where $\Psi(x_\mu)$ is the perturbation of metric. The second concept is the perturbation of inflationary scalar field
\begin{equation}
    \phi(x_\mu)=\bar{\phi}(\tau)+\delta \phi(x_\mu),
    \label{inflaton field}
\end{equation}
where $\bar{\phi}(\tau)$ denotes the background of inflation and $\delta\phi(x_\mu)$ is the inflation perturbation. Consider a universe described by the FRW metric \eqref{frwl metric} and the scalar field $\phi$ of inflation, whose corresponding action is given by \cite{Nojiri:2017ncd}:
\begin{equation}
   \mathcal{S}_{F(R)}=\int d^{4} x \sqrt{-g}\left(\frac{F(R)}{2 \kappa^{2}}-\frac{1}{2} \partial_{\mu} \phi \partial^{\mu} \phi-V(\phi)\right),
\end{equation}
where $\kappa^2$ is related to the gravitational constant via $\kappa^2 = 8\pi G = \frac{1}{M_{\text{p}}^2}$, with $M_{\text{p}}$ denoting the reduced Planck mass. By replacing the scalar curvature $R$ in the Einstein–Hilbert action and introducing a suitable function of the scalar curvature \cite{Nojiri:2006ri}, one obtains
\begin{equation}\label{SEH}
   \mathcal{S}_{EH}=\int d^{4} x \sqrt{-g}\left(\frac{R}{2 \kappa^{2}}-\frac{1}{2} \partial_{\mu} \phi \partial^{\mu} \phi-V(\phi)\right),
\end{equation}
the first term $\frac{R}{2 \kappa^{2}}$ represents the gravitational component and is independent of $\phi$, the second term $\frac{1}{2} \partial_{\mu} \phi \partial^{\mu} \phi$ represents the kinetic energy, and the third term $V(\phi)$ represents the potential energy.The quadratic action for the physical scalar perturbation is obtained by expanding about the background and eliminating the constraint variables 
\begin{equation}
     \mathcal{S}_{EH}=\int d\eta d^{3} x a^4\left( -\frac{1}{2}({\phi}')^{2}+\frac{1}{2}(\partial_{i} \phi)^{2} +\frac{R}{2 \kappa^{2}}-V(\phi)\right).
     \label{standard canonical action}
\end{equation}
The potential is included in the Mukhanov-Sasaki variable via the background evolution.The perturbative expansion is taken about the time-dependent background
\begin{equation}\label{potential energy}
    V(\phi)=\alpha \mu^{2}\left(1-\mathrm{e}^{-\sqrt{\frac{2}{3}} \phi}\right)^{2}\sim \frac{2}{3}\alpha \mu^{2},
\end{equation}
where the dimensionless scaling factor $\alpha$ is typically taken to be one. The parameter $\mu^2$ denotes the energy density, which determines the magnitude of the gravitational effect produced by the scalar field $\chi$ at different locations.The potential normalization is a model parameter and must be specified independently of the reduced Planck mass. From this, we obtain the quadratic action
\begin{equation}
     \mathcal{S}^{(2)}=\frac{1}{2}\int d\eta d^{3} x\big( -\pi^{2}+(\partial_{i} v)^{2} -\frac{8}{3}\alpha\mu^2a^2v^2\big).
\end{equation}
Since $F(R)$/$R$ here does not contain the scalar field $\phi$, and therefore does not contain the field function $\hat{v} (\eta,\vec{x})$ either, it makes no contribution whatsoever to the expansion of the quadratic action $\mathcal{S}^{(2)}$. In order to construct the Hamiltonian operator, therefore one could define the canonical momentum
\begin{equation}
\pi(\eta,\vec{x})=\frac{\delta L}{\delta{v}'(\eta,\vec{x})} ={v}'-\frac{{z}'}{z}v,
\end{equation}
where $z=\sqrt{2\epsilon}a$, $\epsilon=-\frac{\dot{H}}{H^2}=1-\frac{{\mathcal{H}}'}{\mathcal{H}^2}$. It will lead to the Hamiltonian by means of $ H=\int d^3xd\eta(\pi v'-\mathcal{L} )$, 
\begin{equation}  \label{hamilton}
H= \frac{1}{2}\int d\eta d^3x  \left [ -{\pi}^2-(\partial _i v)^2+\frac{ {z}'}{z}(\pi v+v\pi)+\frac{8}{3}\alpha\mu^2a^2v^2\right ],
\end{equation}
Using the Fourier decomposition as follows, 
\begin{equation}\label{v}
\hat{v} (\eta,\vec{x})=\int \frac{d^3k}{(2\pi)^{3/2}} \sqrt{\frac{1}{2k}}(\hat{c }_{-\vec{k} }^{\dagger}v_{\vec {k}}^{\ast }(\eta )e^{-i\vec{k\cdot }\vec{x}}+ {c_{\vec k}}v_{\vec k}e^{i\vec{k\cdot }\vec{x}}),
\end{equation}
\begin{equation}\label{pi}
\hat{\pi} (\eta,\vec{x})=i\int \frac{d^3k}{(2\pi)^{3/2}}\sqrt{\frac{k}{2}}(\hat{c }_{-\vec{k}}^{\dagger}u_{\vec k}^{\ast }(\eta )e^{-i\vec{k\cdot }\vec{x}}-\hat{c}_{\vec k}u_{\vec k}e^{i\vec{k\cdot }\vec{x}}),
\end{equation}
 where $\hat{c }_{-\vec{k} }^{\dagger}$ and $\hat{c}_{\vec k}$ represent the creation and annihilation operators, respectively. Being armed with these two Fourier modes, the Hamiltonian \eqref{hamilton} will become as follows 
\begin{equation}\label{standard hamilton}
\begin{split}
\hat{H}=\int{d^{3}k}\hat{H}_{k}=&\int{d^{3}k}\bigg[(A-k)\big[\hat{c }_{-\vec{k} }^{\dagger}\hat{c}_{-\vec{k}}+\hat{c}_{\vec k}\hat{c}_{\vec k}^{\dagger }\big]+(A+i\frac{z{}'}{z})\hat{c}_{\vec k}^{\dagger }\hat{c}_{-\vec{k} }^{\dagger }+(A-i\frac{z{}'}{z})\hat{c}_{\vec k}\hat{c}_{-\vec{k} }\bigg].
\end{split}
\end{equation}
where, $A=\frac{4\alpha\mu^2a^2}{3k}$. For these two modes, we have $[\hat c_k,\hat c_{-k}]=[\hat c_k,\hat c^\dagger_{-k}]=[\hat c^\dagger_{k},\hat c_{-k}]=[\hat c^\dagger_{k},\hat c^\dagger_{-k}]=0$, $[\hat c_k,\hat{c}_k^\dagger]=[\hat c_{-k},\hat{c}_{-k}^\dagger]=1$. 

For obtaining its wave function in Fock space, we need the following unitary operator 
\begin{equation}
\mathcal{U}_{k}=  \hat{\mathcal{S}}_{k}(r_{k},\varphi_{k})\hat{\mathcal{R}}_{k}(\theta_{k}),
\label{unitary operator}
\end{equation}
with 
\begin{equation}
\hat{\mathcal{R} }_{\vec{k}}(\theta _{k})=\exp[-i\theta _{k(\eta)}(\hat{c}_{\vec {k}}\hat{c}_{\vec{k}}^{\dagger }+\hat{c}_{-\vec{ k}}^{\dagger }\hat{c}_{-\vec{ k}})],
\label{rotation operator}
\end{equation}
\begin{equation}
\hat{\mathcal{S}}_{\vec{k}}(r_{k},\varphi_{k})=\exp[r_{k}(\eta)(e^{-2i\varphi_{k}(\eta)}\hat{c}_{\vec{k}}\hat{c}_{-\vec{k}}-e^{2i\varphi_{k}(\eta)}\hat{c}^{\dagger}_{-\vec{k}}\hat{c}^{\dagger}_{\vec{k}})],
\label{squeezed operator}
\end{equation}
where $\hat{\mathcal{R}}_{\vec{k}}(\theta_{k})$ is the rotation operator and $\hat{\mathcal{S}}_{\vec{k}}(r_{k},\varphi_{k})$ is the squeezed operator.For the chosen vacuum input and the stated ordering of the squeezed and rotation operators, the rightmost rotation contributes only a global phase. Applying the squeezed operator $\hat{\mathcal{S}}_{\vec{k}}(r_{k},\varphi_{k})$ to the vacuum state yields,
\begin{equation}\label{wave function of squeezed state}
\left | \psi  \right \rangle_{sq} =\frac{1}{\cosh r_k}\sum_{n=0}^{\infty }(-1)^{n}e^{2in\varphi_{k}}\tanh^{n}r_{k} \left | n; n \right \rangle _{\vec{k},-\vec{k}}. 
\end{equation}
Substituting the two-mode squeezed state into the Schrödinger equation with the derived Hamiltonian gives $i\frac{d}{d\eta} \left | \psi  \right \rangle _{sq}=\hat{H}_{k}\left | \psi  \right \rangle_{sq} $, one can obtain the evolution equation of $r_k$ and $\varphi_k$, respectively,
\begin{equation}\label{rk and phik}
{r}'_{k}=A\sin(2\varphi_{k})-\frac{z'}{z} \cos(2\varphi_{k}),\\{\varphi}'_k=k-A+\frac{z' }{z}\sin(2\varphi_{k})\coth(2r_{k})+A\cos(2\varphi_{k})\coth(2r_{k}).
\end{equation}
 If we approximate $\epsilon$ to be a constant, then we can get $\frac{z' }{z} =\frac{a' }{a}$. We can then obtain the differential equations \eqref{rk and phik} for $r_k$ and $\varphi_k$ with respect to $\eta$
\begin{equation}\label{rk of scalar inflation1}
    {r}'_{k}=A\sin{2\varphi_{k}}-\frac{a'}{a}\cos{2\varphi_{k}},
\end{equation}
\begin{equation}\label{phik of scalar inflation}
 \begin{split}
{\varphi}'_k=k-A+\frac{a'}{a}\sin{2\varphi_{k}}\coth{2r_k}+A\cos{2\varphi_{k}}\coth{2r_k},
 \end{split}
 \end{equation}
We use \(y=\log_{10}a\) as the evolution variable, for which \(d/d\eta=(aH/\ln10)d/dy\). The relation \(a=-1/(H_0\eta)\) is used only in the constant-\(H\) approximation.
\begin{equation}
    \frac{d{r}_{k}}{dy}=\frac{\ln{10}}{aH_0}A\sin{2\varphi_{k}}-\ln{10}\cos{2\varphi_{k}}, 
\label{rk_evolution}
\end{equation}
\begin{equation}
    \frac{d{\varphi}_{k}}{dy}=\frac{\ln{10}}{aH_0}(k-A)+\ln{10}\sin{2\varphi_{k}}\coth{2r_k}+\frac{\ln{10}}{aH_0}A\cos{2\varphi_{k}}\coth{2r_k}.
\end{equation}
 Once the variable for the squeezed parameter $r_k$ and $\varphi_k$ has been changed from theconformal time $\eta$ to $y$, a plot of the evolution of the squeezed parameter can be obtained. Here, $A$ fully accounts for the contribution of the potential energy $V(\phi)$. After we consider the slow-roll condition, the following evolution equation for squeezed parameter $r_k$ and $\varphi_k$ can be obtained
\begin{equation}\label{last rk and phik of canonical}
  \begin{split}
       & \quad\quad\quad\frac{d{r}_{k}}{dy}=-\ln{10}\cos{2\varphi_{k}}, \\&\frac{d{\varphi}_{k}}{dy}=\frac{\ln{10}}{aH_0}k+\ln{10}\sin{2\varphi_{k}}\coth{2r_k}.
  \end{split}
\end{equation}
The comoving wavenumber \(k\) sets the gradient scale. The potential still affects the mode evolution through the background quantities \(H\), \(\epsilon\), and \(z\).

\subsection{$f(\phi, R)$ inflation}
\label{Inﬂation in theories of gravity}

In this section, we investigate inflationary models of the $f(\phi,R)$ type. In order to retain the explicit effects of the nonminimal coupling function $f(\phi,R)$ in the complexity analysis, we work in the Jordan frame. This choice is particularly convenient because,A conformal-frame change reorganizes the background and canonical variables. We therefore state the frame and normalization used for all state diagnostics. In the Jordan frame, by contrast, the contribution of $f(\phi,R)$ can be kept manifest at the level of the quadratic action. A similar strategy was adopted in Ref. \cite{Zhang:2022bde}, where the Jordan frame description was used to address the non-propagation problem of adiabatic modes in mimetic gravity. As a consistency check, the choice $f(\phi,R)=R$ reduces smoothly to the standard canonical scenario described by Eq.~\eqref{standard canonical action}. A general class of $f(\phi,R)$ inflationary theories is described by the following action \cite{Noh:2004bc,Hwang:2002fp,Hwang:2001fb,Hwang:2000jh,Hwang:1996xh,Sebastiani:2015kfa,Farajollahi:2011odw},
\begin{equation}\label{action of theories of gravity}
    \mathcal{S}_{f(\phi,R)}=\int d^{4} x \sqrt{-g}\left(\frac{f( \phi,R)}{2 \kappa^{2}}-\frac{1}{2} \omega(\phi) g^{\mu \nu} \partial_{\mu} \phi \partial_{\nu} \phi-V(\phi)\right),
\end{equation}
where $f(\phi, R)$ is a smooth function of the scalar field $\phi$ and the Ricci scalar $R$. Here, $R=12H^{2}+\dot{H}$, and $\kappa^{2}$ is chosen such that $\frac{1}{M_{p}^{2}}$. Note that if the kinetic coupling satisfies $\omega(\phi) \neq 1$, the scalar field is non-canonical. The action \eqref{action of theories of gravity} represents a generalized framework. Specific models such as pure $f(\phi, R)$ gravity, canonical and non-canonical scalar-tensor theories, and non-minimally coupled models of the form $f(\phi, R) = f(\phi)R$ can all be recovered as special cases. The three terms on the right-hand side of Eq. \eqref{action of theories of gravity}—$\frac{f(\phi,R)}{2 \kappa^{2}}$, $-\frac{1}{2} \omega(\phi) g^{\mu \nu} \partial_{\mu} \phi \partial_{\nu} \phi$, and $-V(\phi)$—correspond to the modified gravity sector, the kinetic energy, and the potential energy of the scalar field $\phi$, respectively. 

The perturbation spectrum must be calculated for the specified coupling and background solution. As an example, we shall consider the theory of the function $f( \phi)$, which possesses a special symmetry $\beta\to \frac{1}{\beta}$, where $\beta>1$. Potentials of this type have already been considered in Ref. \cite{Odintsov:2016jwr}. Therefore, assuming that the function $f( \phi)$ equal to
\begin{equation}\label{fR1}
    f(\phi)=\frac{1+\xi\left(\mathrm{e}^{-\beta n \phi}+\mathrm{e}^{-\frac{1}{\beta} n \phi}\right)}{2},
\end{equation}
where the parameters $\beta$ and $n$ are positive real parameters and $\frac{n}{\beta}\ge1$. The choice of the scalar potential $V(\phi)$ should ensure that it does not affect the dynamics compared to the function $f( \phi)$. For example, it may include higher powers of the exponential term appearing in Eq. \ref{fR1}. In the present case, we may simply choose $V(\phi) = \Lambda$, where $\Lambda$ is a normalized parameter. Of course, we could also choose the same expression as in Eq. \ref{potential energy}.The full coupling in Eq. \ref{fR1} is invariant under \(\beta\leftrightarrow\beta^{-1}\). The subsequent one-exponential approximation is restricted to the stated asymptotic regime. For simplicity, we choose a system of physical units such that $\kappa^2 = 1$; furthermore, we choose $\xi = 1$ to simplify the intermediate expressions. The function $f( \phi)$ can be approximated as
\begin{equation}\label{fR2}
    f(\phi) \simeq \frac{1+\mathrm{e}^{-\frac{n}{\beta} \phi}}{2}.
\end{equation}
As can be seen from Eq. \ref{fR2}, in contrast to canonical scalar field inﬂation, the modified gravity in inflation that incorporates modified gravitational effects includes the scalar field $\phi$. To simplify the complexity involved in correcting for gravity when dealing with the action \ref{action of theories of gravity}, we shall expand the correction for gravity to the second order. After introducing the field function $\phi (x_\mu )=\phi_0(\eta )+\delta \phi (x_\mu )$, combining all second-order terms containing $v$ yields the quadratic action
\begin{equation}
     \mathcal{S}^{(2)}=\int d\tau d^{3} x \left( -(\pi)^{2}-(\partial_{i} v)^{2} +\big(B-A\big)v^{2}\right).
\end{equation}
where $B=\frac{n^2a^{2}R}{\kappa^{2}\beta^2}$. Similarly, by substituting Eqs. \eqref{v} and \eqref{pi} into the definition of the Hamiltonian, $H = \int d^3x d\eta (\pi v' - \mathcal{L})$, we can express the corresponding Hamiltonian in terms of the creation and annihilation operators as follows
\begin{equation}\label{H of FR}
    \begin{split}
       \hat{H}=\int{d^{3}k}\hat{H}_{k}=\int{d^{3}k}&\bigg[(A-B-k)\big[\hat{c }_{-\vec{k} }^{\dagger}\hat{c}_{-\vec{k}}+\hat{c}_{\vec k}\hat{c}_{\vec k}^{\dagger }\big]+(A-B+i\frac{a{}'}{a})\hat{c}_{\vec k}^{\dagger }\hat{c}_{-\vec{k} }^{\dagger }\\&+(A-B-i\frac{a{}'}{a})\hat{c}_{\vec k}\hat{c}_{-\vec{k} }\bigg].
    \end{split}
\end{equation}
We then solve the Schrödinger equation $i\frac{d}{d\eta} \left | \psi  \right \rangle _{sq}=\hat{H}_{k}\left | \psi  \right \rangle_{sq} $ using the squeezed wave function \ref{wave function of squeezed state}, and the evolution equations for the squeezed parameters $r_k$ and $\varphi_k$ can be derived as follows
\begin{equation}\label{rk in fR}
   {r}'_{k}=(A-B)\sin{2\varphi_{k}}-\frac{a{}'}{a}\cos{2\varphi_{k}},
\end{equation}
\begin{equation}\label{phik in fR}
\begin{split}
      {\varphi}'_{k}=B+k-A+\frac{a{}'}{a}\sin{2\varphi_{k}}\coth2{r}_{k}+(A-B)\cos{{2\varphi_{k}}}\coth{{2r_{k}}}.
\end{split}
\end{equation}
Similarly, we need to replace the variable $\eta$ with $y=\log_{10}a$ in Eqs. \ref{rk in fR} and \ref{phik in fR}.
\begin{equation}
    \frac{d{r}_{k}}{dy}=\frac{\ln{10}}{aH_0}(A-B)\sin{2\varphi_{k}}-\ln{10}\cos{2\varphi_{k}}, 
\end{equation}
\begin{equation}
    \frac{d{\varphi}_{k}}{dy}=\frac{\ln{10}}{aH_0}(B+k-A)+\ln{10}\sin{2\varphi_{k}}\coth2{r}_{k}+(A-B)\frac{\ln{10}}{aH_0}\cos{{2\varphi_{k}}}\coth{{2r_{k}}}.
\end{equation}
In Eqs. \ref{rk in fR} and \ref{phik in fR}, $A$ represents the contribution from potential energy, whilst $B$ represents the contribution from the modified gravity $f(\phi,R)$. In other words, the squeezed parameter $r_k$ and $\varphi_k$ evolves under the combined influence of the potential energy and the modified gravity. When the contribution from the modified gravity $B$ is neglected, we can revert to Eqs. \ref{rk of scalar inflation1} and \ref{phik of scalar inflation}.Dropping an explicit potential-dependent term is an additional approximation and must be justified by its size relative to the retained terms
\begin{equation}\label{last rk and phik of fR}
    \begin{split}
       & \quad\quad\quad\quad\quad\quad\frac{d{r}_{k}}{dy}=-\frac{\ln{10}}{aH_0}B\sin{2\varphi_{k}}-\ln{10}\cos{2\varphi_{k}},\\&\frac{d{\varphi}_{k}}{dy}=\frac{\ln{10}}{aH_0}(B+k)+\ln{10}\sin{2\varphi_{k}}\coth2{r}_{k}-\frac{\ln{10}}{aH_0}B\cos{{2\varphi_{k}}}\coth{{2r_{k}}}.
    \end{split}
\end{equation}
Consequently, the following figures shows two curves of squeezed parameters in canonical scalar field inflation and $f(\phi,R)$ inflation.

\begin{figure}[h!]
    \centering
    \includegraphics[width=0.75\linewidth]{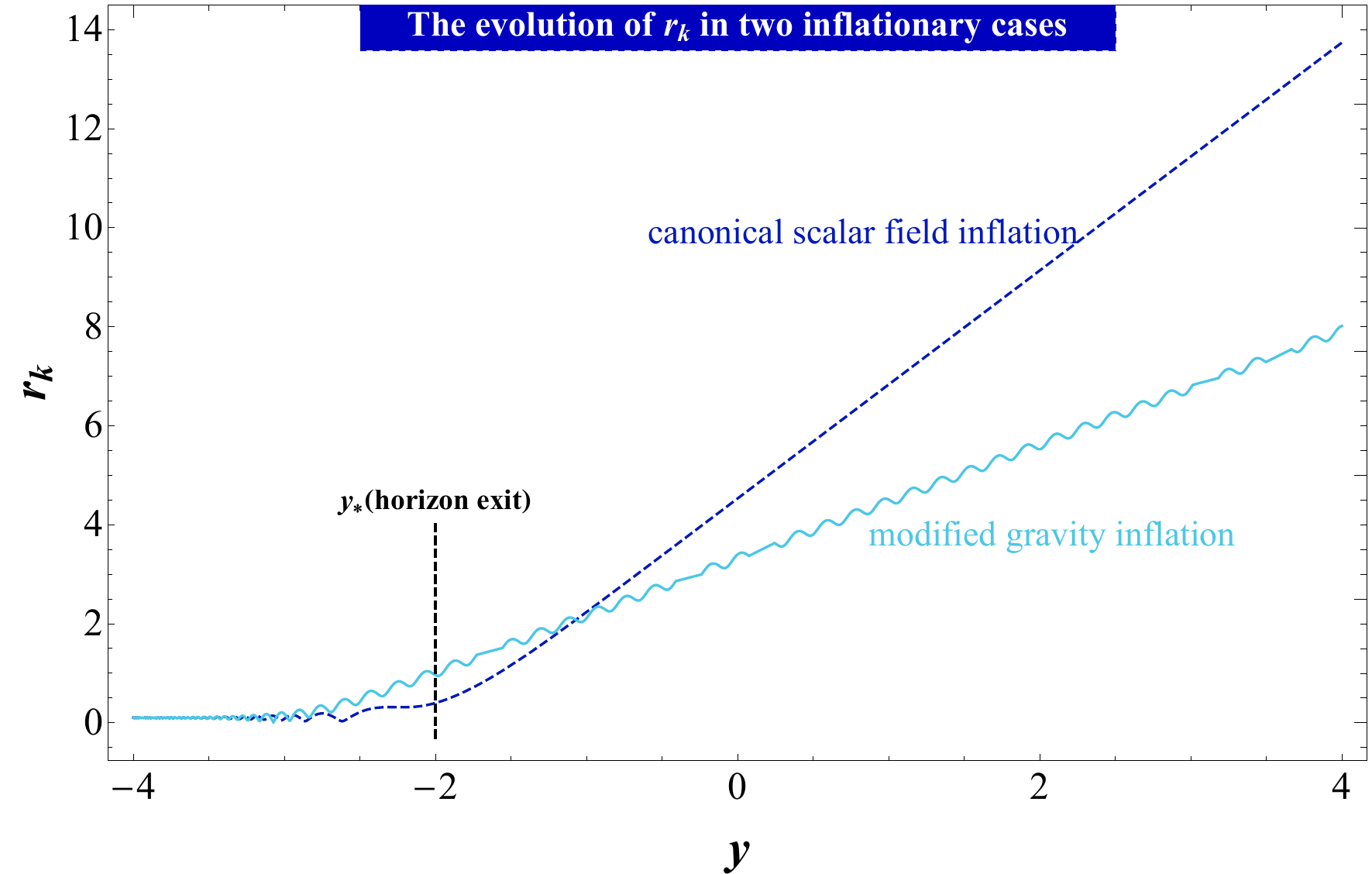}
    \caption{The numerical solutions of $r_k$ in terms of $y=\log_{10}a$ in the canonical scalar field inflation and $f(\phi, R)$ inflation. Our plots adopt $H_0=1$, $k=0.01$ and $\alpha\mu^2=\kappa^2=1$. And we set $\frac{n}{\beta}\ge1$, the horizon exit at $y_*=-2$.}
    \label{fig:rk}
\end{figure}
\begin{figure}[h!]
    \centering
    \includegraphics[width=0.75\linewidth]{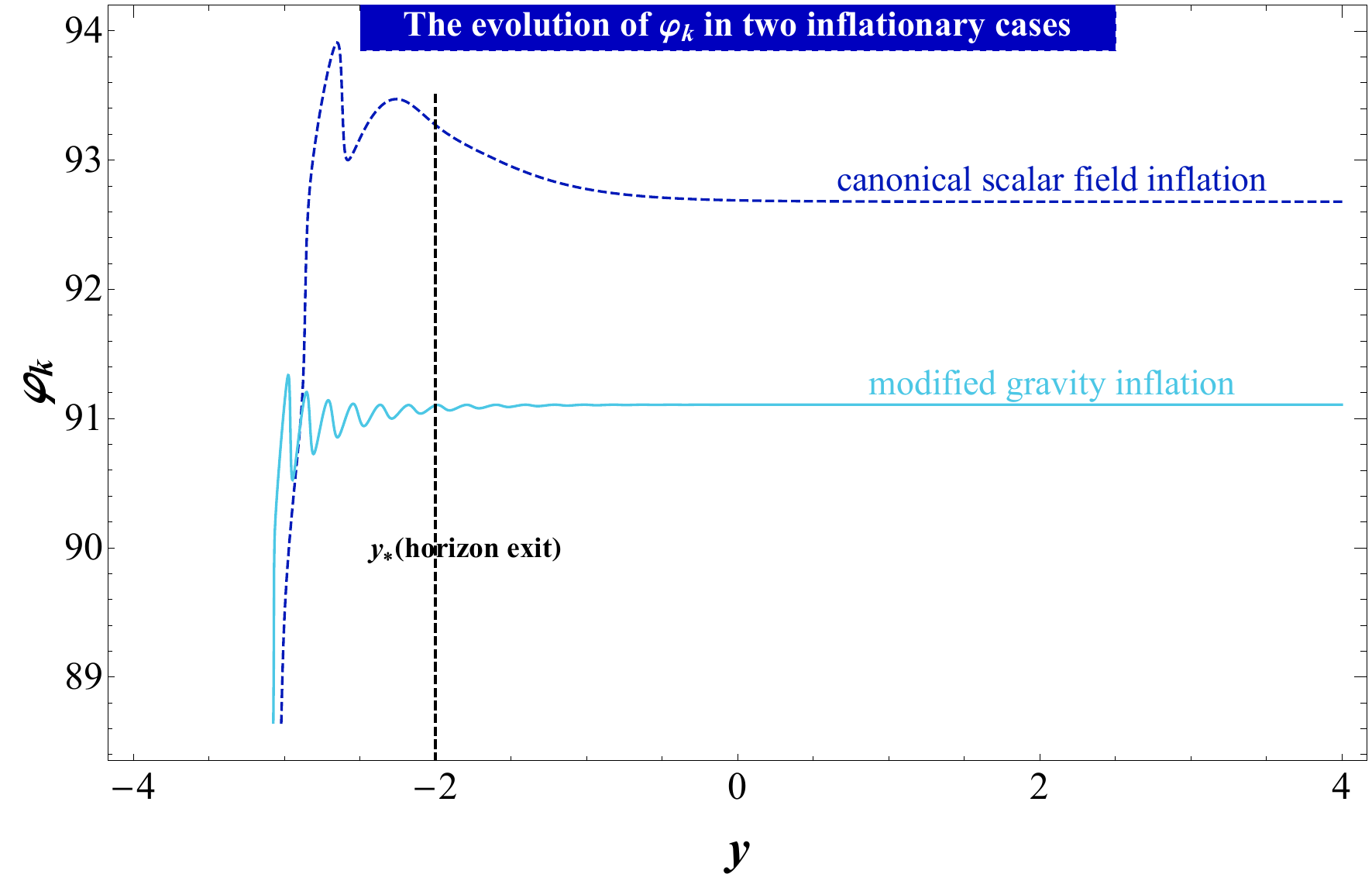}
    \caption{The numerical solutions of $\varphi_k$ in terms of $y=\log_{10}a$ in the canonical scalar field inﬂation and $f(\phi, R)$ inflation. Our plots adopt $H_0=1$, $k=0.01$ and $\alpha\mu^2=\kappa^2=1$. And we set $\frac{n}{\beta}\ge1$, the horizon exit at $y_*=-2$.}
    \label{fig:phik}
\end{figure}

Fig. \ref{fig:rk} illustrates the evolution of the squeezed parameter $r_k$ in canonical scalar field inflation and $f(\phi, R)$ inflation, and took into account their evolution following the introduction of slow-roll conditions.Relative to canonical scalar field inflation, the \(f(\phi,R)\) model modifies both the magnitude and the phase structure of the primordial two-mode squeezed state. after horizon exit, the canonical squeezed strength \(r_k\) grows approximately linearly, whereas the modified gravity result exhibits a slower average growth accompanied by persistent small oscillations. Since the mean occupation number of each mode is given by \(\bar n_k=\sinh^2r_k\), the reduced value of \(r_k\) indicates a suppressed average pair-production rate for the \((\vec{k},-\vec{k})\) modes. In the strong-squeezed regime, \(\bar n_k\simeq e^{2r_k}/4\), so a difference of only a few units in \(r_k\) can correspond to a substantial difference in the occupation number. The modified gravity coupling therefore moderates the accumulation of squeezed rather than eliminating it. 

The squeezed angle also approaches a different late-time value in Fig. \ref{fig:phik}. This difference should be interpreted as a change in the asymptotic orientation of the covariance matrix and in the field–momentum correlations, rather than simply as a reduction in the squeezed strength. The modified gravity contribution remains present in the coupled evolution equations after horizon exit and continues to affect the squeezed angle through the effective pump term. The residual phase evolution then modulates the growth rate of \(r_k\), producing the oscillatory structure observed in the modified gravity curve.

These changes have direct consequences for complexity-based diagnostics. In the adopted Krylov representation, \(K=\sinh^2r_k\), so the reduced growth and oscillatory behavior of \(r_k\) are inherited by the Krylov complexity. This indicates a slower and phase-modulated spreading of the primordial perturbation state in Krylov space. The circuit complexity is additionally sensitive to the squeezed angle through its nonlinear dependence on \(r_k\) and \(\varphi_k\), and therefore provides information about the changing orientation of the Gaussian-state covariance matrix. Overall, the \(f(\phi,R)\) coupling reorganizes the pair-production, phase-space rotation, and information-spreading dynamics of primordial perturbations. These results reveal a a model-dependent modification of the quantum dynamics of modified gravity.  

To summarize Sec. \ref{Squeezed parameter in the inﬂationary paradigm},we compare the evolution of the squeezed parameter \(r_k\) and the squeezed angle \(\varphi_k\) in canonical scalar field inflation and \(f(\phi,R)\) inflation. The results show that modified gravity slows the growth of \(r_k\) and induces small-amplitude oscillations. These oscillations mainly arise because the modified gravity coupling changes the effective frequency and effective pump term of the perturbation modes, leading to persistent phase evolution of the squeezed angle and enhanced interference between the positive- and negative-frequency components. Consequently, the growth of \(r_k\) becomes phase-modulated and develops an oscillatory structure. Meanwhile, the squeezed angle approaches different late-time values in the two models. This indicates that modified gravity changes not only the squeezed strength but also the asymptotic orientation of the primordial two-mode squeezed state in phase space and the correlations between the field and its conjugate momentum. Since \(K=\sinh^2r_k\), the oscillatory behavior of the squeezed parameter is inherited by the Krylov complexity, whereas the circuit complexity depends on both \(r_k\) and \(\varphi_k\). Therefore, the \(f(\phi,R)\) coupling jointly modifies the pair-production process, phase-space rotation, covariance-matrix orientation, and information spreading of primordial perturbations. Whether these effects further influence the curvature power spectrum, non-Gaussianity, or primordial black-hole formation remains to be investigated.

\section{Circuit complexity and Krylov complexity }
\label{evolution of Circuit complexity and Krylov complexity}

In this section, we compute the Krylov complexity following the closed system approach of Ref. \cite{Adhikari:2022oxr}, alongside the circuit complexity following Ref. \cite{Nielsen:2005mkt}. Before analyzing these complexities within canonical scalar field inflation and $f(\phi, R)$ inflation, we briefly review the fundamental kinematics of inflation to establish the necessary groundwork for our subsequent numerical investigations. As previously discussed, the action \eqref{action of theories of gravity} describes single field inflation. Before the hot Big Bang, the universe underwent a period of exponential expansion (typically from $10^{-36}$ to $10^{-32}$ seconds), parameterized by the scale factor $a(\eta)$ in the metric \eqref{frwl metric}. For specific details, please refer to Refs. \cite{Liu:2021nzx,Li:2021kfq}.

\subsection{Circuit complexity of canonical scalar field inflation and $f(\phi, R)$ inﬂation }
\label{Circuit complexity of inflation}

In this section, we employ the geometric approach proposed by Nielsen et al. \cite{Nielsen:2005mkt,Nielsen:2006cea,Dowling:2006tnk} to investigate the circuit complexity in both canonical scalar field inflation and $f(\phi, R)$ inflation.We evaluate the circuit complexity using the numerical solutions for \(r_k\) and \(\varphi_k\), with the reference state and admissible circuit complexity in Sec. \ref{Circuit complexity}.

\begin{figure}[h!]
    \centering
    \includegraphics[width=0.75\linewidth]{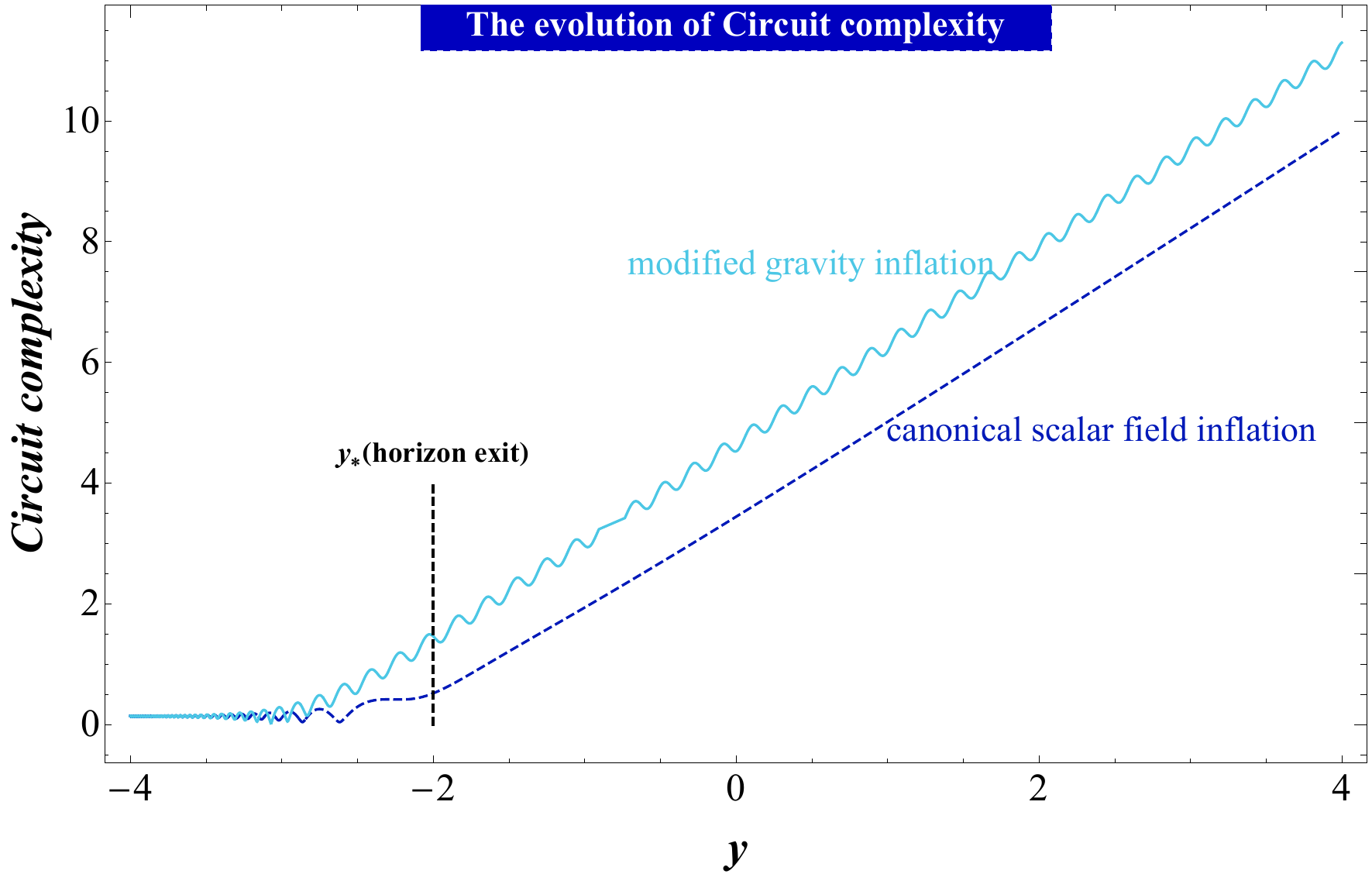}
    \caption{The numerical solutions of circuit complexity in the canonical scalar field inflation and $f(\phi,R)$ inﬂation in terms of $y=\log_{10}a$. Our plots
adopt $H_0 = 1$, $k=0.01$, $\frac{n}{\beta}\ge1$, $\alpha\mu^2=\kappa^2=1$ and the horizon exit at $y_*=-2$.The Lyapunov exponent can be identified from the linear growth rate of circuit complexity \cite{Ali:2019zcj}.}
    \label{fig:circuit complexity}
\end{figure}

Figure \ref{fig:circuit complexity} illustrates the numerical evolution of circuit complexity in canonical scalar field inflation and $f(\phi, R)$ inflation, evaluated using the numerical solutions for the squeezed parameters $r_k$ and $\varphi_k$.Before horizon exit,two circuit complexity curves exhibit erratic oscillations, and the introduction of modified gravity results in even greater amplitudes. The values of circuit complexityafter horizon exit are significantly higher in the $f(\phi, R)$ inflation.It is worth noting that the Lyapunov exponent in \(f(\phi,R)\) inflation is almost identical to that in the canonical case, which suggests that modified gravity hardly changes the degree of quantum chaoticity in the early universe. The persistent modulation of the circuit complexity after horizon exit originates from the coupled evolution of the squeezed strength and the squeezed angle in the modified gravity background. In the present model, the decrease of \(k/(aH_0)\) after horizon exit does not necessarily eliminate the modified gravity contribution. Consequently, the squeezed angle continues to be affected by the effective pump term, and its residual phase variation further modulates the growth of the squeezed strength \(r_k\). Since the circuit complexity is a nonlinear function of \(r_k\) and \(\varphi_k\), even small oscillations in the orientation of the covariance matrix can produce visible oscillations in the circuit complexity, although the squeezed angle appears to approach an approximately constant value. This result indicates that the shortest path connecting the reference state and the target state on the Gaussian-state manifold is modulated in a phase-dependent manner. Accordingly, the phenomenon shown in the figure should be interpreted as an irregular modulation of the circuit complexity induced by the modified gravity inflation. Fundamentally, circuit complexity quantifies the minimum number of unitary gates required to evolve a given reference state (the primordial vacuum state) into a target state (the quantum state at the end of inflation). As clearly demonstrated in Fig. \ref{fig:circuit complexity}, the final complexity value in the $f(\phi, R)$ model ($\sim 13$) substantially exceeds that of canonical inflation ($\sim 10$).For the specified circuit complexity prescription, the modified gravity target state has a larger preparation cost relative to the common reference state. 

To summarize Sec. \ref{Circuit complexity of inflation}, we have mapped the numerical evolution of circuit complexity for both canonical scalar field and $f(\phi, R)$ inflation.As the foundational baseline framework, canonical scalar field inflation possesses a simple evolutionary trajectory. When the modified gravity coupling \(f(\phi,R)\) is introduced, the circuit complexity undergoes rapid oscillations after horizon exit. Combining the Lyapunov exponent with the definition of circuit complexity, we demonstrate that modified gravity does not alter the quantum chaoticity during the inflation. Quantitatively, the number of unitary operators required to construct the final target state is significantly larger in $f(\phi, R)$ inflation than in the canonical case.The modified gravity coupling continues to affect the coupled evolution of the squeezed strength and squeezed angle, thereby phase-modulating the shortest state-preparation path on the Gaussian-state manifold and causing the circuit complexity to exhibit irregular oscillations even after horizon exit.

\subsection{Krylov complexity of canonical scalar field inflation and $f(\phi, R)$ inﬂation}
\label{Krylov complexity of inflation}

Previous methodologies for calculating Krylov complexity, such as the standard framework established in Ref. \cite{Parker:2018yvk}, are fundamentally rooted in static, flat space-times. However, cosmic inflation drives an exponential expansion of the dynamical background. Concurrently, the inflationary process is accompanied by quantum decoherence—such as the minimal decoherence scenario discussed in Ref. \cite{Burgess:2022nwu}—which introduces information loss.For this reason, in this chapter we introduce systematic dissipation to investigate information loss in modified gravity inflation. The calculation below concerns unitary Gaussian evolution of a mode pair. Decoherence requires an additional open system description.

The definition of Krylov complexity was derived in Sec. \ref{Krylov complexity}
\begin{equation}
 \begin{split}
 K=\sum_{n=1}^{k_n}n\left | \Psi_{n} \right | ^{2},
 \end{split}\label{Krylov complexity of close system}
 \end{equation}
where $\Psi_{n}$ corresponds to the operator wave function in the two-mode squeezed state \ref{wave function of squeezed state}, which is expressed as follows
\begin{equation}\label{operator wave function}
\Psi_{n} =\frac{1}{\cosh r_k}(-1)^ne^{2in\varphi_{k}}\tanh^{n}r_{k}. 
\end{equation}
Combining Eqs. \ref{Krylov complexity of close system} and \ref{operator wave function} and applying rule $\sum_{m=0}^{\infty}=z/(1-z)^2$ yields an expression for Krylov complexity
\begin{equation}\label{Krylov complexity of inflation 1}
   K=\sum_{n} n\left|\Psi_{n}\right|^{2}=\sum_{n=0}^{\infty} n \frac{\tanh ^{2 n} r_{k}}{\cosh ^{2} r_{k}}=\sinh ^{2} r_{k}
\end{equation}
It is clear that the Krylov complexity exclusively reflects the squeezed parameter $r_k$. Having established the precise definitions of the Krylov complexity, the Lanczos coefficients $b_n$, and the dissipative term $c_n$, we are now positioned to explore the underlying physics within various inflationary models. These coefficients are derived from the recursion relation by utilizing the tridiagonal form of the Liouvillian superoperator
\begin{equation}\label{Liouvillian superoperator}
    \mathcal{L}|\mathcal{O}_n) = -i c_n |\mathcal{O}_n) + b_{n+1} |\mathcal{O}_{n+1}) + b_n |\mathcal{O}_{n-1})
\end{equation}
where the Liouvillian superoperator $\mathcal{L}$ corresponds to the Hamiltonian operator $\hat{H}_k$.

\subsubsection{Lanczos coefficients and dissipative term}
To proceed, one only needs the corresponding operator wave function—or the squeezed strength $r_k$—along with the Hamiltonian $\hat{H}_k$. In canonical scalar field inflation, these equations are expressed as 
\begin{equation}\label{standard hamilton}
\begin{split}
\hat{H}=\int{d^{3}k}\hat{H}_{k}=&\int{d^{3}k}\bigg[(A-k)\big[\hat{c }_{-\vec{k} }^{\dagger}\hat{c}_{-\vec{k}}+\hat{c}_{\vec k}\hat{c}_{\vec k}^{\dagger }\big]+(A+i\frac{a{}'}{a})\hat{c}_{\vec k}^{\dagger }\hat{c}_{-\vec{k} }^{\dagger }+(A-i\frac{a{}'}{a})\hat{c}_{\vec k}\hat{c}_{-\vec{k} }\bigg].
\end{split}
\end{equation}
We separate the Hamiltonian into pair-creation and pair-annihilation terms and a number-diagonal term.
\begin{equation}
    \hat{H}_{\text{close}}=(A+i\frac{a{}'}{a})\hat{c}_{\vec k}^{\dagger }\hat{c}_{-\vec{k} }^{\dagger }+(A-i\frac{a{}'}{a})\hat{c}_{\vec k}\hat{c}_{-\vec{k} },
\end{equation}
\begin{equation}
\begin{split}
\hat{H}_{\text{open}}=(A-k)\big[\hat{c }_{-\vec{k} }^{\dagger}\hat{c}_{-\vec{k}}+\hat{c}_{\vec k}\hat{c}_{\vec k}^{\dagger }\big].
\end{split}
\end{equation}
Applying Hamiltonian to the Krylov basis $|\mathcal{O}_{n})$ or the two mode squeezed state $|n ; n\rangle_{\vec{k},-\vec{k}}$ yields the Lanczos coefficients $b_n$ and dissipative term $c_n$
\begin{equation}
\begin{split}
 \hat{H}_{\text{close}} |\mathcal{O}_{n})=(n+1)\left[\frac{4\alpha\mu^2a^2}{3k}+i\frac{a{}'}{a}\right] |\mathcal{O}_{n+1} )+n\left[\frac{4\alpha\mu^2a^2}{3k}-i\frac{a{}'}{a}\right] |\mathcal{O}_{n-1}),
\end{split}
\end{equation}
\begin{equation}
    \left.\hat{H}_{\text{open}}|\mathcal{O}_{n})=\left.(2n+1)\left[\frac{4\alpha\mu^2a^2}{3k}-k\right] \right\rvert\, \mathcal{O}_{n}\right).
\end{equation}
Based on the detailed definition of the Lanczos coefficient in Ref. \cite{Muck:2022xfc} and dissipative term in Ref \cite{Li:2024kfm}, we obtain
\begin{equation}\label{b and c in scalar inflation}
    b_{n}=n \sqrt{\left(\frac{4\alpha\mu^2a^2}{3k}\right)^{2}+\left(\frac{a^{\prime}}{a}\right)^{2}},~  ~  c_n=i(2n+1)\bigg(\frac{4\alpha\mu^2a^2}{3k}\bigg),
\end{equation}
where the Lanczos coefficients scale as $b_n \propto n$, characterizing an idealized closed system \cite{Li:2024kfm, Parker:2018yvk}. Concurrently, under the open system framework,a real number-diagonal term contributes to unitary evolution and does not, by itself, define environmental dissipation. Following the framework established in Ref. \cite{Bhattacharya:2022gbz}, our investigation clearly demonstrates that the quadratic action of the single field inflation decomposes into distinct closed system and open system contributions. 

We will now use the same theoretical approach as above to discuss the Lanczos coefficients $b_n$ and dissipative term $c_n$ when $f(\phi,R)$ is introduced during theinflationary period. The corresponding Hamiltonian \ref{H of FR} is as follows
\begin{equation}
    \begin{split}
        \hat{H}=\int{d^{3}k}\hat{H}_{k}=\int{d^{3}k}&\bigg[(A-B-k)\big[\hat{c }_{-\vec{k} }^{\dagger}\hat{c}_{-\vec{k}}+\hat{c}_{\vec k}\hat{c}_{\vec k}^{\dagger }\big]+(A-B+i\frac{a{}'}{a})\hat{c}_{\vec k}^{\dagger }\hat{c}_{-\vec{k} }^{\dagger }\\&+(A-B-i\frac{a{}'}{a})\hat{c}_{\vec k}\hat{c}_{-\vec{k} }\bigg].
    \end{split}
\end{equation}
Similarly, it can be broken down into an ideal closed system and an open system 
\begin{equation}
     \hat{H}_{\text{close}}=(A-B+i\frac{a{}'}{a})\hat{c}_{\vec k}^{\dagger }\hat{c}_{-\vec{k} }^{\dagger }+(A-B-i\frac{a{}'}{a})\hat{c}_{\vec k}\hat{c}_{-\vec{k} }\bigg],
\end{equation}
\begin{equation}
    \hat{H}_{\text{open}}=\bigg[(A-B-k)\big[\hat{c }_{-\vec{k} }^{\dagger}\hat{c}_{-\vec{k}}\bigg].
\end{equation}
Applying Hamiltonian to the Krylov basis $|\mathcal{O}_{n})$
\begin{equation}
    \left.\left.\hat{H}_{\text{close}} |\mathcal{O}_{n}) \left.=(n+1)\left[(A-B+i\frac{a{}'}{a})\right] \right\rvert\, \mathcal{O}_{n+1}\right) \left.-n\left[(A-B-i\frac{a{}'}{a})\right] \right\rvert\, \mathcal{O}_{n-1}\right),
\end{equation}
\begin{equation}
    \hat{H}_{\text{open}}|\mathcal{O}_{n}) \left.=\left.(2n+1)\left[(A-B-k)\right] \right\rvert\, \mathcal{O}_{n}\right),
\end{equation}
Combining this with $\mathcal{L}$ \ref{Liouvillian superoperator}, we obtain
\begin{equation}\label{b and c in fr inflation}
\begin{split}
    & b_{n}=n \sqrt{\bigg(\frac{4\alpha\mu^2a^2}{3k}-\frac{n^2a^{2}R}{\kappa^{2}\beta^2}\bigg)^2+\left(\frac{a^{\prime}}{a}\right)^{2}}, \\&
    c_n=i(2n+1)\bigg(\frac{4\alpha\mu^2a^2}{3k}-\frac{n^2a^{2}R}{\kappa^{2}\beta^2}-k\bigg).
\end{split}
\end{equation}
We can then obtain the statistical properties of Lanczos coefficients $b_n$ and dissipative term $c_n$ during the inflationary phase of the early universe.
\begin{figure}[h!]
    \centering
    \includegraphics[width=0.5\linewidth]{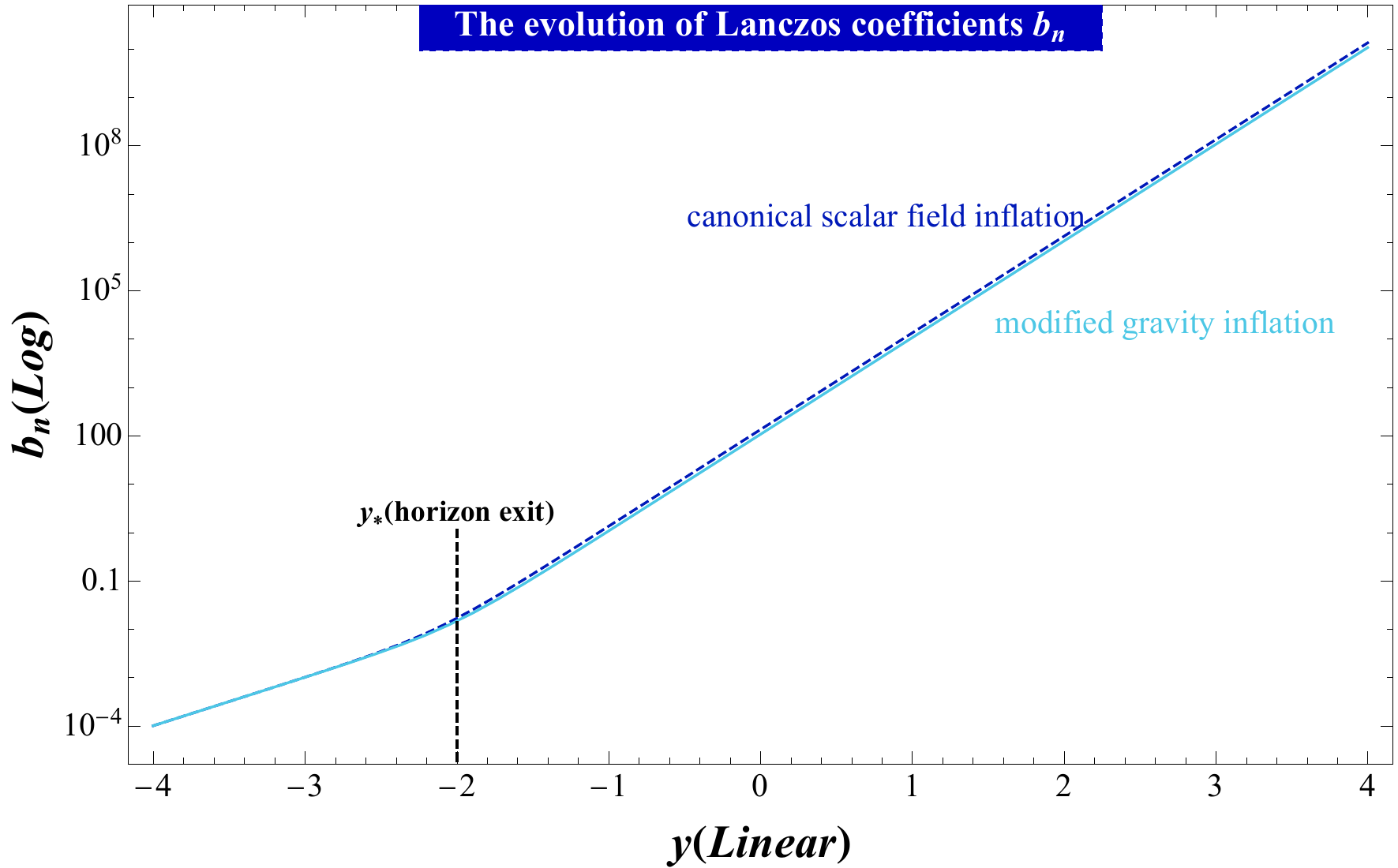}\includegraphics[width=0.5\linewidth]{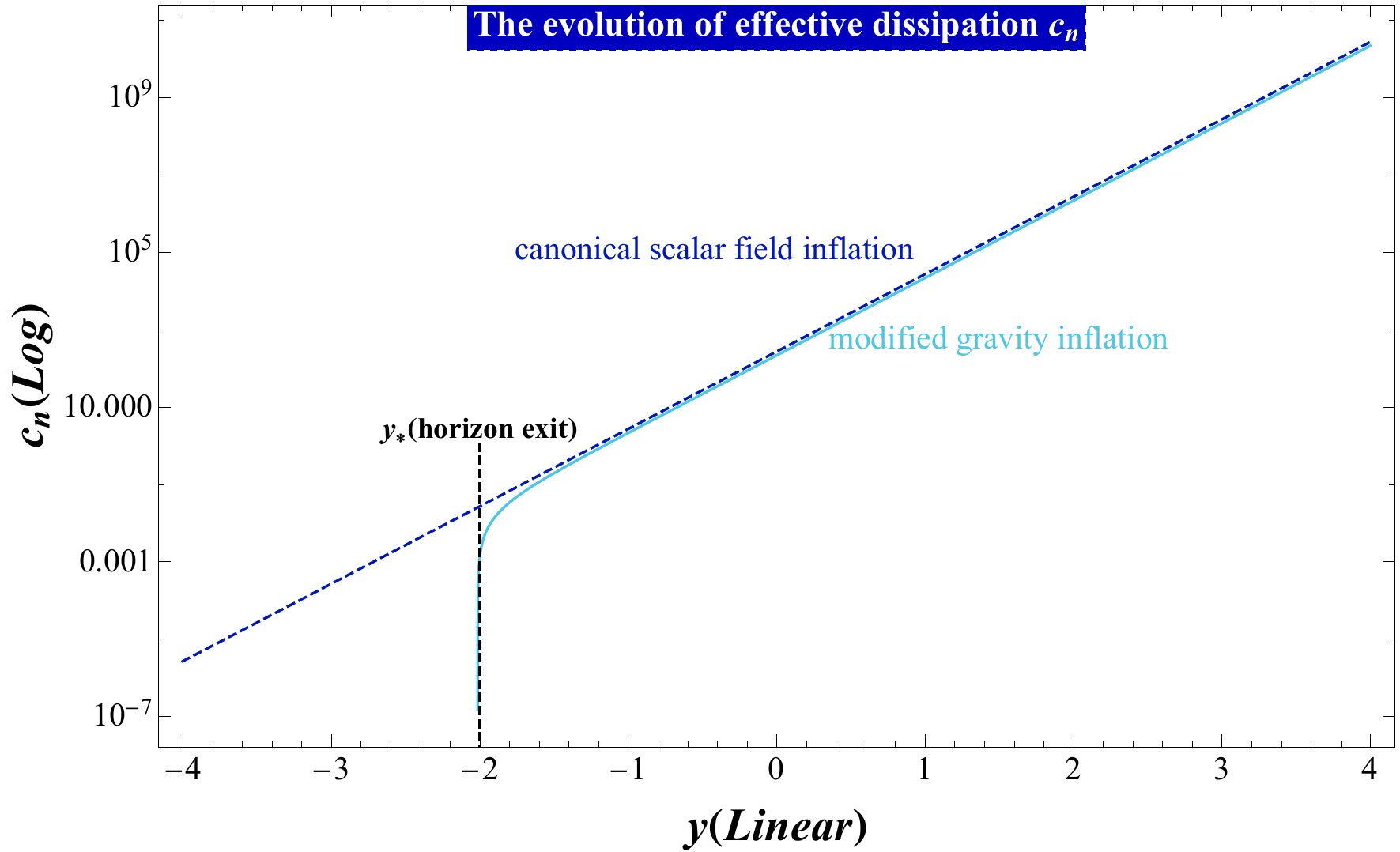}
    \caption{The numerical solutions of $b_n$ and $c_n$ in terms of $y=\log_{10}a$ in the canonical scalar field and $f(\phi,R)$ inflation. As the values are very large when only the modified gravity $f(\phi, R)$ is taken into account, we have $\log$-scaled the vertical axis to improve visual clarity. Our plots
adopt $H_0 = 1$, $k=0.01$, $\frac{n}{\beta}\ge1$, $\alpha\mu^2=\kappa^2=1$ and the horizon exit at $y_*=-2$. }
    \label{fig:bn of scalar inflation and fR inflation}
\end{figure}

Fig. \ref{fig:bn of scalar inflation and fR inflation} illustrates the numerical evolution of the Lanczos coefficients $b_n$ and $c_n$ during the inflationary epoch.The Lanczos coefficient \(b_n\) characterizes the spreading of operators in Krylov space and can also characterize the quantum chaoticity of the system. Its growth provides information about operator dynamics. It is evident that the selected diagnostics show no significant change in the effective quantum chaoticity for the parameters considered during the inflationary epoch, which is consistent with the conclusion previously drawn from the analysis of the Lyapunov exponents in Fig. \ref{fig:circuit complexity}.

When the universe is treated as an open system that exchanges information with its environment, the corresponding dissipative effects must be properly taken into account. This effect is described by the dissipation term \(c_n\) shown in Fig. \ref{fig:bn of scalar inflation and fR inflation}, which quantifies the efficiency of information exchange between the system and its surroundings. After incorporating the \(f(\phi,R)\) modified gravity effects, the dissipation rate of the system is nearly zero before horizon exit, indicating that information loss during this stage is negligible. After horizon exit, the dissipation rate is nearly identical to that in canonical inflation, indicating that modified gravity has little effect on the exchange of information between the system and its environment during the later stage of inflation. 

\subsubsection{Krylov complexity and Krylov entropy}
\label{Krylov complexity and Krylov entropy}

The Krylov complexity captures operator growth, and its definition \ref{Krylov complexity of inflation 1} in the two-mode squeezed state has been established
\begin{equation}\label{K}
    K=\sinh^2{r_k},
\end{equation}
it is directly related only to squeezed strength $ r_k$. In the canonical scalar field inflation and $f(\phi,R)$ inflation, the evolution equations for the squeezed strength $r_k$ can beobtained form Eqs. \ref{last rk and phik of canonical} and \ref{last rk and phik of fR}
\begin{equation}\label{rk without vphi of canonical}
    \frac{d{r}_{k}}{dy}=-\ln{10}\cos{2\varphi_{k}},
\end{equation}
\begin{equation}\label{rk without vphi of fR}
    \frac{d{r}_{k}}{dy}=-\frac{\ln{10}}{aH_0}B\sin{2\varphi_{k}}-\ln{10}\cos{2\varphi_{k}}.
\end{equation}
Substituting the numerical solutions of Eqs. \ref{rk without vphi of canonical},  \ref{rk without vphi of fR} and consider the slow-roll conditions corresponding to the two types of inflation into Eq. \ref{K} yields the following evolution of the Krylov complexity.
\begin{figure}[h!]
    \centering
    \includegraphics[width=0.75\linewidth]{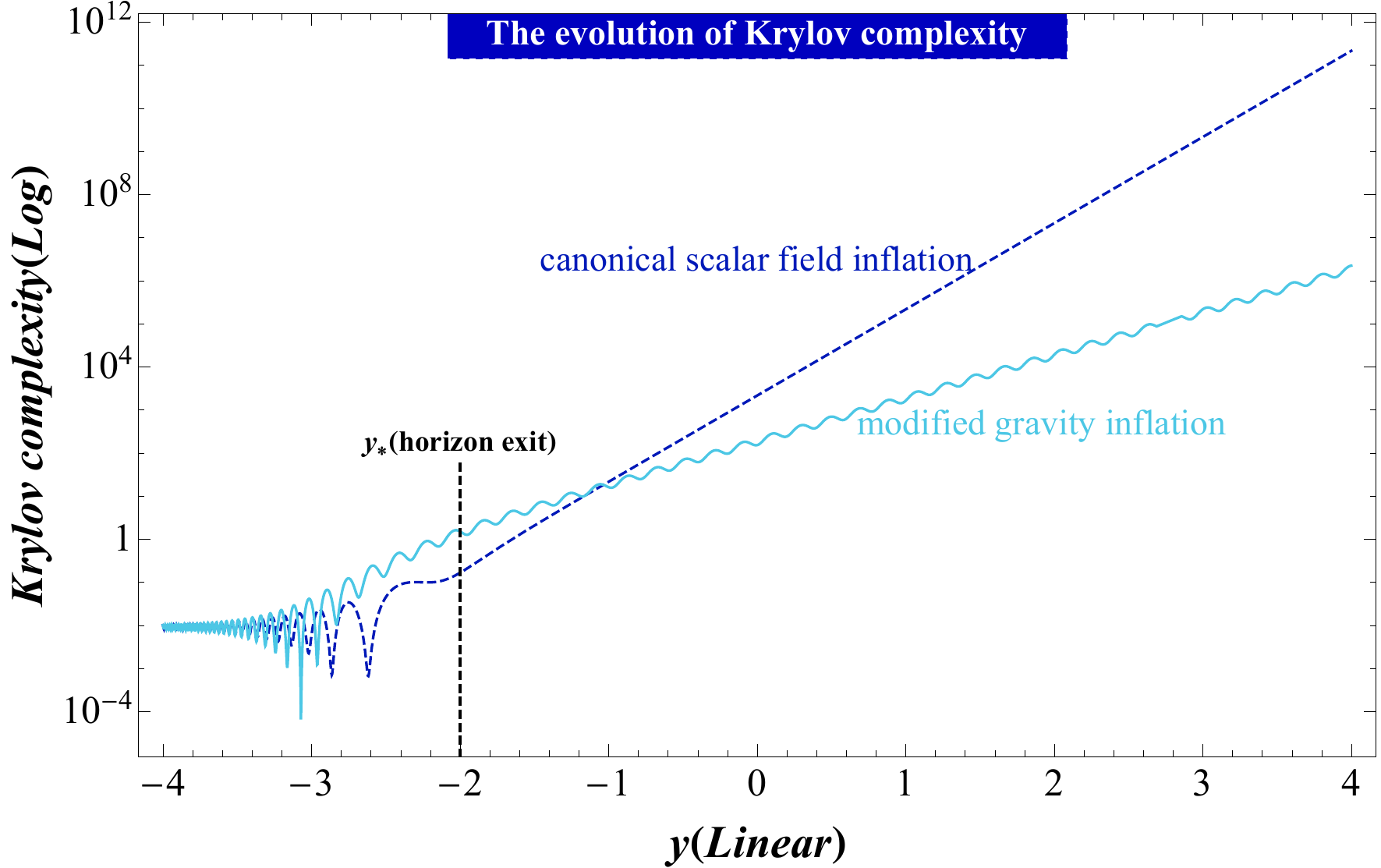}
    \caption{The numerical solutions of Krylov complexity in terms of $y=\log_{10}a$ in the canonical scalar field and $f(\phi,R)$ inflation. We have $\log$-scaled the vertical axis to improve visual clarity. Our plots
adopt $H_0 = 1$, $k=0.01$, $\frac{n}{\beta}\ge1$, $\alpha\mu^2=\kappa^2=1$, and the horizon exit at $y_*=-2$. }
    \label{fig:Krylov complexity of scalar inflation and fR inflation}
\end{figure}

Fig. \ref{fig:Krylov complexity of scalar inflation and fR inflation} shows the evolution of Krylov complexity in the canonical scalar field inflation and $f(\phi,R)$ inflation. In both inflationary models,the numerical curve is approximately consistent with exponential growth over the displayed interval. Curvature perturbations are defined both inside and outside the Hubble radius. Their state evolves as the background changes. In the context of modern cosmology, the growth in operators captured by Krylov complexity corresponds to the curvature perturbation operator. Consequently, Krylov complexity increases rapidly in an exponential manner after horizon exit.At matched late times, the canonical curve has a larger pair-basis spread than the modified gravity curve. This is because the squeezed strength $r_k$ driving the evolution of Krylov complexity differs vastly between the two models. However, this enormous discrepancy arises precisely due to the effects of $f(\phi,R)$ .The difference in pair-basis spread follows algebraically from the difference in squeezed strength. Its relation to the curvature power spectrum requires the mode normalization and squeezed phase. In the adopted Krylov basis, the Krylov complexity of the two-mode squeezed state is \(K=\sinh^2 r_k\). Therefore, after horizon exit oscillations of Krylov complexity are inherited directly from the oscillatory evolution of the squeezed strength. The modified gravity contribution continues to affect the effective pump term and the squeezed angle dynamics after horizon exit, thereby modulating the pair-production rate and producing oscillations in the mean Krylov position. This effect indicates a phase-dependent modification of the information spreading of primordial perturbations in Krylov space. Furthermore, discusses an issue that is relevant to in Ref. \cite{Adhikari:2022oxr}.In the chosen Fock representation, the spread over the pair-number basis equals the mean occupation number of either mode: $\left\langle\hat{n}_{k}\right\rangle=\left\langle\hat{n}_{-k}\right\rangle=K$. The modified gravity model exhibits a smaller late-time Krylov complexity than the canonical model, together with persistent oscillations after horizon exit. These features indicate that modified gravity suppresses and periodically modulates the pair-production process of the \((\vec{k},-\vec{k})\) modes. This identity refers to a basis-dependent occupation number and does not represent a flux of particles into a spatial region. This not only explains the conjecture that the Krylov complexity equals the average number of particles in each mode, but also accounts for the quantum fluctuations—the source of curvature perturbations—as the seeds for the formation of large-scale cosmic structures.

Furthermore, Krylov entropy has already been defined in Ref. \cite{Barbon:2019wsy}
\begin{equation}\label{K entropy}
    \begin{split}
        S_{K} & =-\sum_{n=0}^{\infty}\left|\Psi_{n}\right|^{2} \ln \left|\Psi_{n}\right|^{2}=-\sum_{n=0}^{\infty} \frac{\tanh ^{2 n} r_{k}}{\cosh ^{2} r_{k}} \ln \frac{\tanh ^{2 n} r_{k}}{\cosh ^{2} r_{k}} \\
& =(\sinh ^{2} r_{k}+1) \ln \big((\sinh ^{2} r_{k}+1)\big)-\sinh ^{2} r_{k} \ln \left(\sinh ^{2} r_{k}\right),
    \end{split}
\end{equation}
By combining the evolution equations for the squeezed strength $r_k$ in the two inflationary models with Eq. \eqref{K entropy}, we obtain the corresponding evolution of the Krylov entropy.
\begin{figure}[h!]
    \centering
    \includegraphics[width=0.75\linewidth]{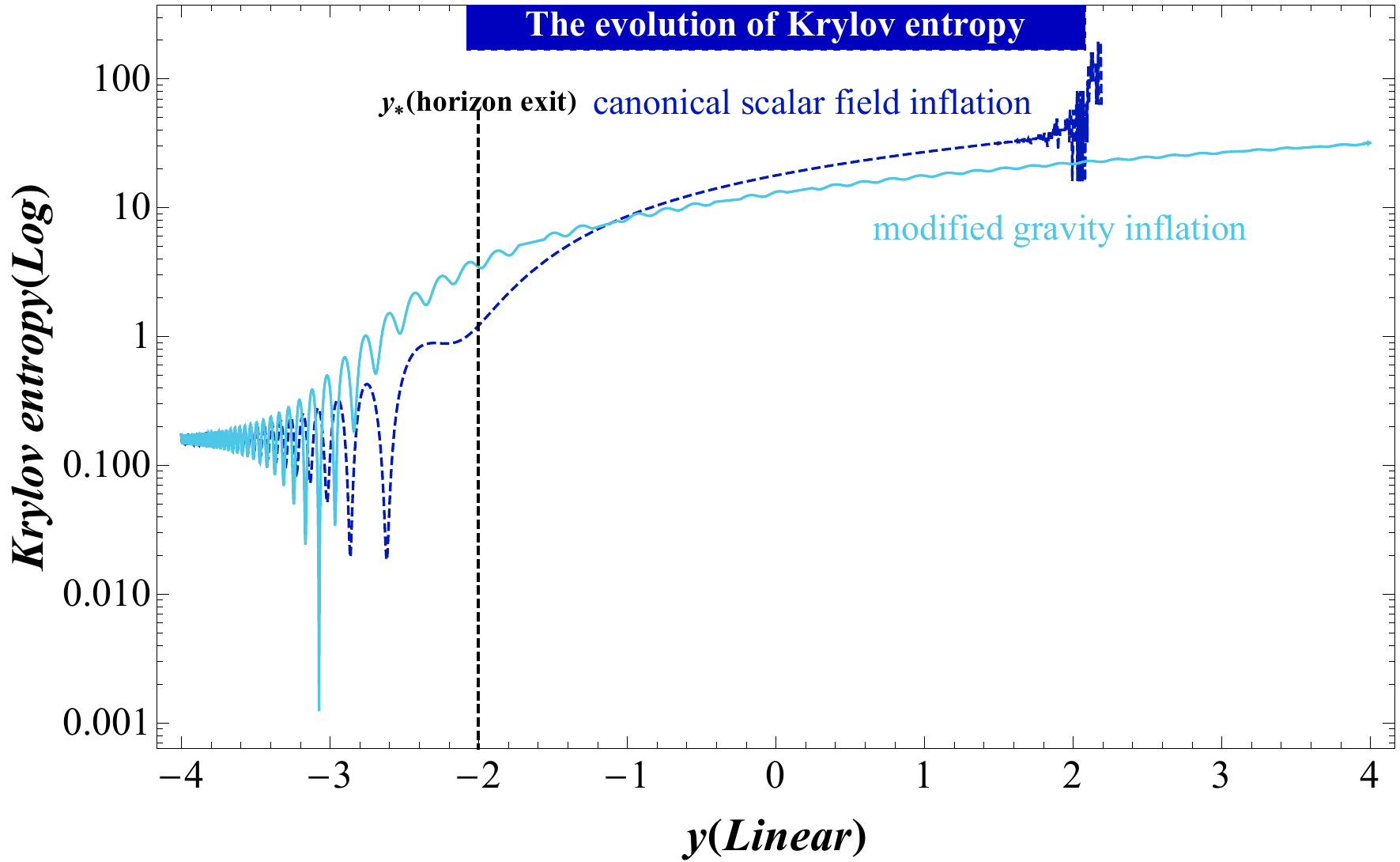}
    \caption{The numerical solutions of Krylov entropy in terms of $y=\log_{10}a$ in the canonical scalar field and $f(\phi,R)$ inflation. We have $\log$-scaled the vertical axis to improve visual clarity. Our plots
adopt $H_0 = 1$, $k=0.01$, $\frac{n}{\beta}\ge1$, $\alpha\mu^2=\kappa^2=1$, and the horizon exit at $y_*=-2$. }
    \label{fig:Krylov entropy of scalar inflation and fR inflation}
\end{figure}

Fig. \ref{fig:Krylov entropy of scalar inflation and fR inflation} shows that the modified gravity model exhibits stronger oscillations before horizon exit and a slower late-time growth of the Krylov entropy than the canonical model. Since \(S_K\) is a function of the squeezed strength through \(\bar{n}_k=\sinh^2r_k\), these features reflect the modified gravity modulation of the pair-production process and the spreading of the primordial perturbation state in Krylov space. after horizon exit, the residual modified gravity pump term continues to affect the evolution of \(r_k\), leading to a slowly varying entropy accompanied by small oscillations. The late-time suppression of \(S_K\) indicates reduced Krylov space spreading within the adopted representation, rather than a reduction of the total thermodynamic entropy of the Universe.

In Section \ref{Krylov complexity of inflation}, we have derived the Krylov complexity, Krylov entropy, Lanczos coefficients $b_n$, and the dissipative terms $c_n$ (with the latter formulated under the cosmic open system framework). The Lyapunov exponent extracted from the linear-growth regime of the circuit complexity, together with the behavior of the Lanczos coefficients $b_n$, indicates that the inclusion of modified gravity has only a negligible effect on the degree of quantum chaoticity during inflation. The dissipative contribution$ c_n$ further suggests that the modified gravity coupling substantially suppresses the dissipation of the perturbation system before horizon exit, while exerting only a minor influence on the exchange of information between the perturbation sector and its environment after horizon exit. After the inclusion of modified gravity, both the Krylov complexity and Krylov entropy exhibit persistent small-amplitude oscillations after horizon exit, while their overall growth rates and late-time values remain lower than those in the canonical scalar field inflation. This indicates that modified gravity modulates the evolution of the squeezed strength, suppressing the mean pair production of opposite-momentum modes and the spreading of the quantum state in Krylov space, while preserving a pronounced time dependence. These results reveal the influence of modified gravity on the quantum statistical dynamics of primordial perturbations.

\section{Summary and Outlook}
\label{conclusion}

In this work, we investigated the circuit complexity and complexity complexity of primordial quantum perturbations in canonical scalar field inflation and \(f(\phi,R)\) modified gravity inflation. The novelty of this work does not lie in introducing a new definition of circuit or Krylov complexity. Rather, it lies in identifying the differential response of several quantum-dynamical diagnostics to the scalar–curvature coupling in \(f(\phi,R)\) inflation. In the adopted Jordan frame model, modified gravity changes the effective pump term and mode frequency, producing persistent oscillations after horizon exit in the squeezed strength $r_k$ and squeezed angle $\varphi_k$. These oscillations are transmitted selectively to the complexity measures: Krylov complexity and Krylov entropy exhibit suppressed and modulated state spreading, whereas the Nielsen circuit complexity is more strongly affected because of its phase sensitivity. At the same time, the Lyapunov and Lanczos diagnostics remain approximately comparable to those of canonical inflation, indicating no significant change in the effective quantum chaoticity for the parameters considered. The open system analysis further shows that modified gravity mainly affects before horizon exit dissipative dynamics, while its relative influence on late-time system–environment information exchange remains limited. Thus, modified gravity does not universally enhance all complexity measures, but reorganizes the amplitude, phase, state-spreading, chaotic, and dissipative responses of primordial quantum perturbations. Below, we elaborate on each point sequentially.

$\bullet$ Our results show that \(f(\phi,R)\) modified gravity changes both the amplitude and phase structure of the primordial two-mode squeezed state. Compared with canonical scalar field inflation, the modified gravity model exhibits more pronounced oscillations in the squeezed parameters before horizon exit. After horizon exit, the squeezed strength \(r_k\) grows more slowly on average and retains persistent small-amplitude oscillations. Since the mean occupation number satisfies \(\bar n_k=\sinh^2 r_k\), the modified gravity model has a smaller mean pair occupation number at late times. Meanwhile, the squeezed angle approaches a different late-time value in the two models, indicating that modified gravity changes the asymptotic orientation of the squeezed state in phase space and the associated field–momentum correlations.

$\bullet$ For circuit complexity, the modified gravity model exhibits stronger oscillations and a larger late-time complexity. Since circuit complexity depends on both the squeezed strength and the squeezed angle, the phase evolution induced by modified gravity modulates the shortest state-preparation path on the Gaussian-state manifold, leading to irregular oscillations in circuit complexity even after horizon exit. For the parameter set considered in this work, the target state in the modified gravity model requires a higher preparation cost. The Lyapunov exponents extracted from the linear-growth regime of circuit complexity are approximately the same in the two models, indicating that the \(f(\phi,R)\) coupling has a limited effect on the effective degree of quantum chaoticity during inflation.

$\bullet$ For Krylov complexity, we obtain, in the selected Fock representation $ K=\sinh^2 r_k $. Thus, Krylov complexity directly inherits the evolution of the squeezed strength. In both models, Krylov complexity grows rapidly after horizon exit. However, the modified gravity model exhibits a slower late-time growth, a smaller final value, and persistent small-amplitude oscillations. Krylov entropy displays a similar behavior, indicating that modified gravity suppresses the spreading of primordial perturbations in the pair-number basis and in Krylov space. These complexity and entropy measures should be understood as diagnostics of quantum-state spreading within the chosen representation, rather than as the total thermodynamic entropy of the Universe or as a spatial particle flux.

$\bullet$ We also investigated the Lanczos coefficients and dissipative contributions within the open system framework. The Lanczos coefficients provide additional information about the quantum chaoticity. Combined with the Lyapunov exponents extracted from circuit complexity, our results indicate that modified gravity does not substantially alter the effective degree of quantum chaoticity. The dissipative contribution shows that modified gravity significantly suppresses dissipation in the perturbation sector before horizon exit, while its influence on the exchange of information between the system and the environment after horizon exit remains relatively small. This suggests that modified gravity affects the unitary squeezed dynamics more strongly than the late-time exchange of information with the environment.

In summary, \(f(\phi,R)\) modified gravity reorganizes the pair-production process, phase-space rotation, quantum-state preparation, Krylov space spreading, and system–environment information exchange of primordial quantum perturbations. And does not significantly alter the effective quantum chaoticity or information loss after horizon exit. Its effect is not simply a universal enhancement or suppression of all complexity. Instead, modified gravity produces a distinct quantum-dynamical pathway by altering the coupled evolution of the squeezed strength and the squeezed angle.

Future work should investigate the dependence of these results on the modified gravity parameters, the initial quantum state, the propagation speed, and the choice of canonical variables. It is also important to connect the evolution of the squeezed parameters and complexity measures with the curvature power spectrum, spectral tilt, tensor-to-scalar ratio, non-Gaussianity, and primordial-black-hole formation. Extending the present analysis to multi-field inflation and developing a more complete reduced-density-matrix and decoherence framework would further clarify the statistical and dynamical properties of primordial quantum perturbations in modified-gravity backgrounds.

\acknowledgments
The research was supported by the National Natural Science Foundation of China under Grant No.12265010, the Project of Guizhou Provincial Department of Science and Technology under Grants No.MS[2025]219 and No.CXTD[2025]030, the Guizhou Provincial Scientific Research Project (Grant No. 2025YJSKYJJ178) and the Scientific Research Project of Guizhou Minzu University (Grant No.GZMUBS7). Finally, we appreciate that Lei-Hua Liu give lots of critical comments for this paper.

\bibliography{Refs}

@article{VanRaamsdonk:2010pw,
    author = "Van Raamsdonk, Mark",
    title = "{Building up spacetime with quantum entanglement}",
    eprint = "1005.3035",
    archivePrefix = "arXiv",
    primaryClass = "hep-th",
    doi = "10.1142/S0218271810018529",
    journal = "Gen. Rel. Grav.",
    volume = "42",
    pages = "2323--2329",
    year = "2010"
}

@article{Caputa:2021sib,
    author = "Caputa, Pawel and Magan, Javier M. and Patramanis, Dimitrios",
    title = "{Geometry of Krylov complexity}",
    eprint = "2109.03824",
    archivePrefix = "arXiv",
    primaryClass = "hep-th",
    doi = "10.1103/PhysRevResearch.4.013041",
    journal = "Phys. Rev. Res.",
    volume = "4",
    number = "1",
    pages = "013041",
    year = "2022"
}

@article{Balasubramanian:2022tpr,
    author = "Balasubramanian, Vijay and Caputa, Pawel and Magan, Javier M. and Wu, Qingyue",
    title = "{Quantum chaos and the complexity of spread of states}",
    eprint = "2202.06957",
    archivePrefix = "arXiv",
    primaryClass = "hep-th",
    doi = "10.1103/PhysRevD.106.046007",
    journal = "Phys. Rev. D",
    volume = "106",
    number = "4",
    pages = "046007",
    year = "2022"
}

@article{Muck:2022xfc,
    author = {M{\"u}ck, Wolfgang and Yang, Yi},
    title = "{Krylov complexity and orthogonal polynomials}",
    eprint = "2205.12815",
    archivePrefix = "arXiv",
    primaryClass = "hep-th",
    doi = "10.1016/j.nuclphysb.2022.115948",
    journal = "Nucl. Phys. B",
    volume = "984",
    pages = "115948",
    year = "2022"
}

@article{Maldacena:2013xja,
    author = "Maldacena, Juan and Susskind, Leonard",
    title = "{Cool horizons for entangled black holes}",
    eprint = "1306.0533",
    archivePrefix = "arXiv",
    primaryClass = "hep-th",
    doi = "10.1002/prop.201300020",
    journal = "Fortsch. Phys.",
    volume = "61",
    pages = "781--811",
    year = "2013"
}

@article{Stanford:2014jda,
    author = "Stanford, Douglas and Susskind, Leonard",
    title = "{Complexity and Shock Wave Geometries}",
    eprint = "1406.2678",
    archivePrefix = "arXiv",
    primaryClass = "hep-th",
    doi = "10.1103/PhysRevD.90.126007",
    journal = "Phys. Rev. D",
    volume = "90",
    number = "12",
    pages = "126007",
    year = "2014"
}

@article{Hartman:2013qma,
    author = "Hartman, Thomas and Maldacena, Juan",
    title = "{Time Evolution of Entanglement Entropy from Black Hole Interiors}",
    eprint = "1303.1080",
    archivePrefix = "arXiv",
    primaryClass = "hep-th",
    doi = "10.1007/JHEP05(2013)014",
    journal = "JHEP",
    volume = "05",
    pages = "014",
    year = "2013"
}

@article{Liu:2013iza,
    author = "Liu, Hong and Suh, S. Josephine",
    title = "{Entanglement Tsunami: Universal Scaling in Holographic Thermalization}",
    eprint = "1305.7244",
    archivePrefix = "arXiv",
    primaryClass = "hep-th",
    reportNumber = "MIT-CTP/4475, MIT-CTP-4475",
    doi = "10.1103/PhysRevLett.112.011601",
    journal = "Phys. Rev. Lett.",
    volume = "112",
    pages = "011601",
    year = "2014"
}

@inproceedings{Aaronson:2016vto,
    author = "Aaronson, Scott",
    title = "{The Complexity of Quantum States and Transformations: From Quantum Money to Black Holes}",
    eprint = "1607.05256",
    archivePrefix = "arXiv",
    primaryClass = "quant-ph",
    month = "7",
    year = "2016"
}

@article{Parker:2018yvk,
    author = "Parker, Daniel E. and Cao, Xiangyu and Avdoshkin, Alexander and Scaffidi, Thomas and Altman, Ehud",
    title = "{A Universal Operator Growth Hypothesis}",
    eprint = "1812.08657",
    archivePrefix = "arXiv",
    primaryClass = "cond-mat.stat-mech",
    doi = "10.1103/PhysRevX.9.041017",
    journal = "Phys. Rev. X",
    volume = "9",
    number = "4",
    pages = "041017",
    year = "2019"
}

@article{Rabinovici:2020ryf,
    author = "Rabinovici, E. and S{\'a}nchez-Garrido, A. and Shir, R. and Sonner, J.",
    title = "{Operator complexity: a journey to the edge of Krylov space}",
    eprint = "2009.01862",
    archivePrefix = "arXiv",
    primaryClass = "hep-th",
    doi = "10.1007/JHEP06(2021)062",
    journal = "JHEP",
    volume = "06",
    pages = "062",
    year = "2021"
}

@article{Jian:2020qpp,
    author = "Jian, Shao-Kai and Swingle, Brian and Xian, Zhuo-Yu",
    title = "{Complexity growth of operators in the SYK model and in JT gravity}",
    eprint = "2008.12274",
    archivePrefix = "arXiv",
    primaryClass = "hep-th",
    doi = "10.1007/JHEP03(2021)014",
    journal = "JHEP",
    volume = "03",
    pages = "014",
    year = "2021"
}

@article{He:2022ryk,
    author = "He, Song and Lau, Pak Hang Chris and Xian, Zhuo-Yu and Zhao, Long",
    title = "{Quantum chaos, scrambling and operator growth in $ T\overline{T} $ deformed SYK models}",
    eprint = "2209.14936",
    archivePrefix = "arXiv",
    primaryClass = "hep-th",
    doi = "10.1007/JHEP12(2022)070",
    journal = "JHEP",
    volume = "12",
    pages = "070",
    year = "2022"
}

@article{Kar:2021nbm,
    author = "Kar, Arjun and Lamprou, Lampros and Rozali, Moshe and Sully, James",
    title = "{Random matrix theory for complexity growth and black hole interiors}",
    eprint = "2106.02046",
    archivePrefix = "arXiv",
    primaryClass = "hep-th",
    doi = "10.1007/JHEP01(2022)016",
    journal = "JHEP",
    volume = "01",
    pages = "016",
    year = "2022"
}

@article{Dymarsky:2019elm,
    author = "Dymarsky, Anatoly and Gorsky, Alexander",
    title = "{Quantum chaos as delocalization in Krylov space}",
    eprint = "1912.12227",
    archivePrefix = "arXiv",
    primaryClass = "cond-mat.stat-mech",
    doi = "10.1103/PhysRevB.102.085137",
    journal = "Phys. Rev. B",
    volume = "102",
    number = "8",
    pages = "085137",
    year = "2020"
}

@article{Camargo:2023eev,
    author = "Camargo, Hugo A. and Jahnke, Viktor and Jeong, Hyun-Sik and Kim, Keun-Young and Nishida, Mitsuhiro",
    title = "{Spectral and Krylov complexity in billiard systems}",
    eprint = "2306.11632",
    archivePrefix = "arXiv",
    primaryClass = "hep-th",
    reportNumber = "IFT-UAM/CSIC-23-61",
    doi = "10.1103/PhysRevD.109.046017",
    journal = "Phys. Rev. D",
    volume = "109",
    number = "4",
    pages = "046017",
    year = "2024"
}

@article{Huh:2023jxt,
    author = "Huh, Kyoung-Bum and Jeong, Hyun-Sik and Pedraza, Juan F.",
    title = "{Spread complexity in saddle-dominated scrambling}",
    eprint = "2312.12593",
    archivePrefix = "arXiv",
    primaryClass = "hep-th",
    reportNumber = "IFT-UAM/CSIC-23-178",
    doi = "10.1007/JHEP05(2024)137",
    journal = "JHEP",
    volume = "05",
    pages = "137",
    year = "2024"
}

@article{Li:2021kfq,
    author = "Li, Ai-chen and Li, Xin-Fei and Zeng, Ding-fang and Liu, Lei-Hua",
    title = "{Cosmological complexity in K-essence}",
    eprint = "2102.12939",
    archivePrefix = "arXiv",
    primaryClass = "gr-qc",
    doi = "10.1016/j.dark.2024.101422",
    journal = "Phys. Dark Univ.",
    volume = "43",
    pages = "101422",
    year = "2024"
}

@article{Li:2023ekd,
    author = "Li, Tao and Liu, Lei-Hua",
    title = "{Cosmological complexity of the modified dispersion relation}",
    eprint = "2309.01595",
    archivePrefix = "arXiv",
    primaryClass = "gr-qc",
    doi = "10.1016/j.physletb.2024.138728",
    journal = "Phys. Lett. B",
    volume = "854",
    pages = "138728",
    year = "2024"
}

@article{Linde:2007fr,
    author = "Linde, Andrei D.",
    title = "{Inflationary Cosmology}",
    eprint = "0705.0164",
    archivePrefix = "arXiv",
    primaryClass = "hep-th",
    doi = "10.1007/978-3-540-74353-8_1",
    journal = "Lect. Notes Phys.",
    volume = "738",
    pages = "1--54",
    year = "2008"
}

@book{Gorbunov:2011zzc,
    author = "Gorbunov, Dmitry S. and Rubakov, Valery A.",
    title = "{Introduction to the theory of the early universe: Cosmological perturbations and inflationary theory}",
    doi = "10.1142/7873",
    year = "2011"
}

@article{Lyth:1998xn,
    author = "Lyth, David H. and Riotto, Antonio",
    title = "{Particle physics models of inflation and the cosmological density perturbation}",
    eprint = "hep-ph/9807278",
    archivePrefix = "arXiv",
    reportNumber = "LANCS-TH-9720, FERMILAB-PUB-97-292-A, CERN-TH-97-383, OUTP-98-39-P",
    doi = "10.1016/S0370-1573(98)00128-8",
    journal = "Phys. Rept.",
    volume = "314",
    pages = "1--146",
    year = "1999"
}

@article{Linde:1983gd,
    author = "Linde, Andrei D.",
    title = "{Chaotic Inflation}",
    doi = "10.1016/0370-2693(83)90837-7",
    journal = "Phys. Lett. B",
    volume = "129",
    pages = "177--181",
    year = "1983"
}

@article{Linde:1993cn,
    author = "Linde, Andrei D.",
    title = "{Hybrid inflation}",
    eprint = "astro-ph/9307002",
    archivePrefix = "arXiv",
    reportNumber = "SU-ITP-93-17",
    doi = "10.1103/PhysRevD.49.748",
    journal = "Phys. Rev. D",
    volume = "49",
    pages = "748--754",
    year = "1994"
}

@article{Sasaki:1995aw,
    author = "Sasaki, Misao and Stewart, Ewan D.",
    title = "{A General analytic formula for the spectral index of the density perturbations produced during inflation}",
    eprint = "astro-ph/9507001",
    archivePrefix = "arXiv",
    reportNumber = "LANCS-TH-9504, OU-TAP-22",
    doi = "10.1143/PTP.95.71",
    journal = "Prog. Theor. Phys.",
    volume = "95",
    pages = "71--78",
    year = "1996"
}

@article{Turok:2002yq,
    author = "Turok, N.",
    editor = "Dunsby, P. and Ellis, G. and Maartens, Roy",
    title = "{A critical review of inflation}",
    doi = "10.1088/0264-9381/19/13/305",
    journal = "Class. Quant. Grav.",
    volume = "19",
    pages = "3449--3467",
    year = "2002"
}

@article{Mukhanov:1990me,
    author = "Mukhanov, Viatcheslav F. and Feldman, H. A. and Brandenberger, Robert H.",
    title = "{Theory of cosmological perturbations. Part 1. Classical perturbations. Part 2. Quantum theory of perturbations. Part 3. Extensions}",
    reportNumber = "BROWN-HET-796, BROWN-HET-800, BROWN-HET-780",
    doi = "10.1016/0370-1573(92)90044-Z",
    journal = "Phys. Rept.",
    volume = "215",
    pages = "203--333",
    year = "1992"
}

@book{Baumann:2014nda,
    author = "Baumann, Daniel and McAllister, Liam",
    title = "{Inflation and String Theory}",
    eprint = "1404.2601",
    archivePrefix = "arXiv",
    primaryClass = "hep-th",
    doi = "10.1017/CBO9781316105733",
    isbn = "978-1-107-08969-3, 978-1-316-23718-2",
    publisher = "Cambridge University Press",
    series = "Cambridge Monographs on Mathematical Physics",
    month = "5",
    year = "2015"
}

@article{Bhattacharya:2022gbz, 
    author = "Bhattacharya, Aranya and Nandy, Pratik and Nath, Pingal Pratyush and Sahu, Himanshu",
    title = "{Operator growth and Krylov construction in dissipative open quantum systems}",
    eprint = "2207.05347",
    archivePrefix = "arXiv",
    primaryClass = "quant-ph",
    doi = "10.1007/JHEP12(2022)081",
    journal = "JHEP",
    volume = "12",
    pages = "081",
    year = "2022"
}

@article{Zhang:2022bde,
    author = "Zhang, Xin-zhe and Liu, Lei-hua and Qiu, Taotao",
    title = "{Mimetic curvaton}",
    eprint = "2207.07873",
    archivePrefix = "arXiv",
    primaryClass = "hep-th",
    doi = "10.1103/PhysRevD.107.043510",
    journal = "Phys. Rev. D",
    volume = "107",
    number = "4",
    pages = "043510",
    year = "2023"
}

@article{Erdmenger:2023wjg,
    author = "Erdmenger, Johanna and Jian, Shao-Kai and Xian, Zhuo-Yu",
    title = "{Universal chaotic dynamics from Krylov space}",
    eprint = "2303.12151",
    archivePrefix = "arXiv",
    primaryClass = "hep-th",
    doi = "10.1007/JHEP08(2023)176",
    journal = "JHEP",
    volume = "08",
    pages = "176",
    year = "2023"
}

@article{Patramanis:2023cwz,
    author = "Patramanis, Dimitrios and Sybesma, Watse",
    title = {{Krylov complexity in a natural basis for the Schr{\"o}dinger algebra}},
    eprint = "2306.03133",
    archivePrefix = "arXiv",
    primaryClass = "quant-ph",
    doi = "10.21468/SciPostPhysCore.7.2.037",
    journal = "SciPost Phys. Core",
    volume = "7",
    pages = "037",
    year = "2024"
}

@article{Fan:2023ohh,
    author = "Fan, Zhong-Ying",
    title = "{Generalised Krylov complexity}",
    eprint = "2306.16118",
    archivePrefix = "arXiv",
    primaryClass = "hep-th",
    month = "6",
    year = "2023"
}

@article{Hashimoto:2023swv,
    author = "Hashimoto, Koji and Murata, Keiju and Tanahashi, Norihiro and Watanabe, Ryota",
    title = "{Krylov complexity and chaos in quantum mechanics}",
    eprint = "2305.16669",
    archivePrefix = "arXiv",
    primaryClass = "hep-th",
    reportNumber = "KUNS-2967",
    doi = "10.1007/JHEP11(2023)040",
    journal = "JHEP",
    volume = "11",
    pages = "040",
    year = "2023"
}

@article{Vasli:2023syq,
    author = "Vasli, M. J. and Babaei Velni, K. and Mohammadi Mozaffar, M. R. and Mollabashi, A. and Alishahiha, M.",
    title = "{Krylov complexity in Lifshitz-type scalar field theories}",
    eprint = "2307.08307",
    archivePrefix = "arXiv",
    primaryClass = "hep-th",
    reportNumber = "IPM/P-2023/53",
    doi = "10.1140/epjc/s10052-024-12609-9",
    journal = "Eur. Phys. J. C",
    volume = "84",
    number = "3",
    pages = "235",
    year = "2024"
}

@article{Bhattacharjee:2022vlt,
    author = "Bhattacharjee, Budhaditya and Cao, Xiangyu and Nandy, Pratik and Pathak, Tanay",
    title = "{Krylov complexity in saddle-dominated scrambling}",
    eprint = "2203.03534",
    archivePrefix = "arXiv",
    primaryClass = "quant-ph",
    doi = "10.1007/JHEP05(2022)174",
    journal = "JHEP",
    volume = "05",
    pages = "174",
    year = "2022"
}

@article{Craps:2025kub,
    author = "Craps, Ben and Pascuzzi, Gabriele and Pedraza, Juan F. and Qu, Le-Chen and Ruan, Shan-Ming",
    title = "{Explicit Connections Between Krylov and Nielsen Complexity}",
    eprint = "2511.15799",
    archivePrefix = "arXiv",
    primaryClass = "hep-th",
    reportNumber = "IFT-UAM/CSIC-25-123",
    month = "11",
    year = "2025"
}

@article{Basu:2025ubf,
    author = "Basu, Pallab and Das, Suman and Nandy, Pratik",
    title = "{Complexity of quadratic quantum chaos}",
    eprint = "2509.04075",
    archivePrefix = "arXiv",
    primaryClass = "hep-th",
    reportNumber = "RIKEN-iTHEMS-Report-25",
    doi = "10.1007/JHEP04(2026)081",
    journal = "JHEP",
    volume = "04",
    pages = "081",
    year = "2026"
}

@article{Domingo:2023kjr,
    author = "Domingo, Laia and Borondo, F. and Scialchi, Gast{\'o}n and Roncaglia, Augusto J. and Carlo, Gabriel G. and Wisniacki, Diego A.",
    title = "{Quantum reservoir complexity by the Krylov evolution approach}",
    eprint = "2310.00790",
    archivePrefix = "arXiv",
    primaryClass = "quant-ph",
    doi = "10.1103/PhysRevA.110.022446",
    journal = "Phys. Rev. A",
    volume = "110",
    number = "2",
    pages = "022446",
    year = "2024"
}

@article{Gill:2023umm,
    author = "Gill, Ankit and Pal, Kunal and Pal, Kuntal and Sarkar, Tapobrata",
    title = "{Complexity in two-point measurement schemes}",
    eprint = "2311.07892",
    archivePrefix = "arXiv",
    primaryClass = "quant-ph",
    doi = "10.1103/PhysRevB.109.104303",
    journal = "Phys. Rev. B",
    volume = "109",
    number = "10",
    pages = "104303",
    year = "2024"
}

@article{Bhattacharjee:2023uwx,
    author = "Bhattacharjee, Budhaditya and Nandy, Pratik and Pathak, Tanay",
    title = "{Operator dynamics in Lindbladian SYK: a Krylov complexity perspective}",
    eprint = "2311.00753",
    archivePrefix = "arXiv",
    primaryClass = "quant-ph",
    reportNumber = "YITP-23-133, RIKEN-iTHEMS-Report-23",
    doi = "10.1007/JHEP01(2024)094",
    journal = "JHEP",
    volume = "01",
    pages = "094",
    year = "2024"
}

@article{Adhikari:2022whf,
    author = "Adhikari, Kiran and Choudhury, Sayantan and Roy, Abhishek",
    title = "{Krylov Complexity in Quantum Field Theory}",
    eprint = "2204.02250",
    archivePrefix = "arXiv",
    primaryClass = "hep-th",
    doi = "10.1016/j.nuclphysb.2023.116263",
    journal = "Nucl. Phys. B",
    volume = "993",
    pages = "116263",
    year = "2023"
}

@article{Jafarizadeh:2006woc,
    author = "Jafarizadeh, M. A. and Salimi, S. and Sufiani, R.",
    title = "{Investigation of continuous-time quantum walk by using Krylov subspace-Lanczos algorithm}",
    eprint = "quant-ph/0606241",
    archivePrefix = "arXiv",
    doi = "10.1140/epjb/e2007-00281-5",
    month = "6",
    year = "2006"
}

@article{Nielsen:2005mkt,
    author = "Nielsen, Michael A.",
    title = "{A geometric approach to quantum circuit lower bounds}",
    eprint = "quant-ph/0502070",
    archivePrefix = "arXiv",
    doi = "10.26421/QIC6.3-2",
    journal = "Quant. Inf. Comput.",
    volume = "6",
    number = "3",
    pages = "213--262",
    year = "2006"
}

@article{Nielsen:2006cea,
    author = "Nielsen, Michael A. and Dowling, Mark R. and Gu, Mile and Doherty, Andrew C.",
    title = "{Quantum Computation as Geometry}",
    eprint = "quant-ph/0603161",
    archivePrefix = "arXiv",
    doi = "10.1126/science.1121541",
    journal = "Science",
    volume = "311",
    number = "5764",
    pages = "1133--1135",
    year = "2006"
}

@article{Dowling:2006tnk,
    author = "Dowling, Mark R. and Nielsen, Michael A.",
    title = "{The geometry of quantum computation}",
    eprint = "quant-ph/0701004",
    archivePrefix = "arXiv",
    doi = "10.26421/QIC8.10-1",
    journal = "Quant. Inf. Comput.",
    volume = "8",
    number = "10",
    pages = "0861--0899",
    year = "2008"
}

@article{Burgess:2022nwu,
    author = "Burgess, C. P. and Holman, R. and Kaplanek, Greg and Martin, Jerome and Vennin, Vincent",
    title = "{Minimal decoherence from inflation}",
    eprint = "2211.11046",
    archivePrefix = "arXiv",
    primaryClass = "hep-th",
    reportNumber = "CERN-TH-2022-174; Imperial/TP/2022/GK/02",
    doi = "10.1088/1475-7516/2023/07/022",
    journal = "JCAP",
    volume = "07",
    pages = "022",
    year = "2023"
}

@article{Guth:1980zm,
    author = "Guth, Alan H.",
    editor = "Fang, Li-Zhi and Ruffini, R.",
    title = "{The Inflationary Universe: A Possible Solution to the Horizon and Flatness Problems}",
    reportNumber = "SLAC-PUB-2576",
    doi = "10.1103/PhysRevD.23.347",
    journal = "Phys. Rev. D",
    volume = "23",
    pages = "347--356",
    year = "1981"
}

@article{Linde:1981mu,
    author = "Linde, Andrei D.",
    editor = "Fang, Li-Zhi and Ruffini, R.",
    title = "{A New Inflationary Universe Scenario: A Possible Solution of the Horizon, Flatness, Homogeneity, Isotropy and Primordial Monopole Problems}",
    reportNumber = "LEBEDEV-81-229",
    doi = "10.1016/0370-2693(82)91219-9",
    journal = "Phys. Lett. B",
    volume = "108",
    pages = "389--393",
    year = "1982"
}

@article{Albrecht:1982wi,
    author = "Albrecht, Andreas and Steinhardt, Paul J.",
    editor = "Fang, Li-Zhi and Ruffini, R.",
    title = "{Cosmology for Grand Unified Theories with Radiatively Induced Symmetry Breaking}",
    reportNumber = "UPR-0185T",
    doi = "10.1103/PhysRevLett.48.1220",
    journal = "Phys. Rev. Lett.",
    volume = "48",
    pages = "1220--1223",
    year = "1982"
}

@article{Guth:1982ec,
    author = "Guth, Alan H. and Pi, S. Y.",
    title = "{Fluctuations in the New Inflationary Universe}",
    doi = "10.1103/PhysRevLett.49.1110",
    journal = "Phys. Rev. Lett.",
    volume = "49",
    pages = "1110--1113",
    year = "1982"
}

@article{Mukhanov:1981xt,
    author = "Mukhanov, Viatcheslav F. and Chibisov, G. V.",
    title = "{Quantum Fluctuations and a Nonsingular Universe}",
    journal = "JETP Lett.",
    volume = "33",
    pages = "532--535",
    year = "1981"
}

@article{Starobinsky:1982ee,
    author = "Starobinsky, Alexei A.",
    title = "{Dynamics of Phase Transition in the New Inflationary Universe Scenario and Generation of Perturbations}",
    doi = "10.1016/0370-2693(82)90541-X",
    journal = "Phys. Lett. B",
    volume = "117",
    pages = "175--178",
    year = "1982"
}

@article{Starobinsky:1979ty,
    author = "Starobinsky, Alexei A.",
    editor = "Khalatnikov, I. M. and Mineev, V. P.",
    title = "{Spectrum of relict gravitational radiation and the early state of the universe}",
    journal = "JETP Lett.",
    volume = "30",
    pages = "682--685",
    year = "1979"
}

@article{Susskind:2014rva,
    author = "Susskind, Leonard",
    title = "{Computational Complexity and Black Hole Horizons}",
    eprint = "1403.5695",
    archivePrefix = "arXiv",
    primaryClass = "hep-th",
    doi = "10.1002/prop.201500092",
    journal = "Fortsch. Phys.",
    volume = "64",
    pages = "24--43",
    year = "2016",
    note = "[Addendum: Fortsch.Phys. 64, 44--48 (2016)]"
}

@article{Susskind:2014moa,
    author = "Susskind, Leonard",
    title = "{Entanglement is not enough}",
    eprint = "1411.0690",
    archivePrefix = "arXiv",
    primaryClass = "hep-th",
    doi = "10.1002/prop.201500095",
    journal = "Fortsch. Phys.",
    volume = "64",
    pages = "49--71",
    year = "2016"
}

@article{Bhattacharyya:2020rpy,
    author = "Bhattacharyya, Arpan and Das, Saurya and Shajidul Haque, S. and Underwood, Bret",
    title = "{Cosmological Complexity}",
    eprint = "2001.08664",
    archivePrefix = "arXiv",
    primaryClass = "hep-th",
    doi = "10.1103/PhysRevD.101.106020",
    journal = "Phys. Rev. D",
    volume = "101",
    number = "10",
    pages = "106020",
    year = "2020"
}

@article{Adhikari:2022oxr,
    author = "Adhikari, Kiran and Choudhury, Sayantan",
    title = "{Cosmological Krylov Complexity}",
    eprint = "2203.14330",
    archivePrefix = "arXiv",
    primaryClass = "hep-th",
    doi = "10.1002/prop.202200126",
    journal = "Fortsch. Phys.",
    volume = "70",
    number = "12",
    pages = "2200126",
    year = "2022"
}

@article{Grishchuk:1974ny,
    author = "Grishchuk, L. P.",
    title = "{Amplification of gravitational waves in an isotropic universe}",
    journal = "Sov. Phys. JETP",
    volume = "40",
    number = "3",
    pages = "409--415",
    year = "1975"
}

@article{Hawking:1982cz,
    author = "Hawking, S. W.",
    title = "{The Development of Irregularities in a Single Bubble Inflationary Universe}",
    reportNumber = "Print-83-0015 (CAMBRIDGE)",
    doi = "10.1016/0370-2693(82)90373-2",
    journal = "Phys. Lett. B",
    volume = "115",
    pages = "295",
    year = "1982"
}

@article{Liu:2021nzx,
    author = "Liu, Lei-Hua and Li, Ai-Chen",
    title = "{Complexity of non-trivial sound speed in inflation}",
    eprint = "2102.12014",
    archivePrefix = "arXiv",
    primaryClass = "gr-qc",
    doi = "10.1016/j.dark.2022.101123",
    journal = "Phys. Dark Univ.",
    volume = "37",
    pages = "101123",
    year = "2022"
}

@article{Ali:2019zcj,
    author = "Ali, Tibra and Bhattacharyya, Arpan and Haque, S. Shajidul and Kim, Eugene H. and Moynihan, Nathan and Murugan, Jeff",
    title = "{Chaos and Complexity in Quantum Mechanics}",
    eprint = "1905.13534",
    archivePrefix = "arXiv",
    primaryClass = "hep-th",
    reportNumber = "YITP-19-45",
    doi = "10.1103/PhysRevD.101.026021",
    journal = "Phys. Rev. D",
    volume = "101",
    number = "2",
    pages = "026021",
    year = "2020"
}

@article{Odintsov:2016jwr,
    author = "Odintsov, S. D. and Oikonomou, V. K.",
    title = "{Inverse Symmetric Inflationary Attractors}",
    eprint = "1611.00738",
    archivePrefix = "arXiv",
    primaryClass = "gr-qc",
    doi = "10.1088/1361-6382/aa69a8",
    journal = "Class. Quant. Grav.",
    volume = "34",
    number = "10",
    pages = "105009",
    year = "2017"
}

@article{Farajollahi:2011odw,
    author = "Farajollahi, H. and Setare, M. and Milani, F. and Tayebi, F.",
    title = "{Cosmic Dynamics in $F(R,\phi)$ Gravity}",
    eprint = "1005.2026",
    archivePrefix = "arXiv",
    primaryClass = "physics.gen-ph",
    doi = "10.1007/s10714-011-1148-z",
    journal = "Gen. Rel. Grav.",
    volume = "43",
    pages = "1657--1669",
    year = "2011"
}

@article{Sebastiani:2015kfa,
    author = "Sebastiani, Lorenzo and Myrzakulov, Ratbay",
    title = "{F(R) gravity and inflation}",
    eprint = "1506.05330",
    archivePrefix = "arXiv",
    primaryClass = "gr-qc",
    doi = "10.1142/S0219887815300032",
    journal = "Int. J. Geom. Meth. Mod. Phys.",
    volume = "12",
    number = "9",
    pages = "1530003",
    year = "2015"
}

@article{Hwang:1996xh,
    author = "Hwang, Jai-chan and Noh, Hyerim",
    title = "{Cosmological perturbations in generalized gravity theories}",
    reportNumber = "PRINT-96-116 (KYUNGPOOK)",
    doi = "10.1103/PhysRevD.54.1460",
    journal = "Phys. Rev. D",
    volume = "54",
    pages = "1460--1473",
    year = "1996"
}

@article{Hwang:2000jh,
    author = "Hwang, Jai-chan and Noh, Hyerim",
    title = "{Cosmological perturbations with multiple scalar fields}",
    eprint = "astro-ph/0009268",
    archivePrefix = "arXiv",
    reportNumber = "KNU-2000-9",
    doi = "10.1016/S0370-2693(00)01253-3",
    journal = "Phys. Lett. B",
    volume = "495",
    pages = "277--283",
    year = "2000"
}

@article{Hwang:2001fb,
    author = "Hwang, Jai-chan and Noh, Hyerim",
    title = "{Cosmological perturbations with multiple fluids and fields}",
    eprint = "astro-ph/0103244",
    archivePrefix = "arXiv",
    doi = "10.1088/0264-9381/19/3/308",
    journal = "Class. Quant. Grav.",
    volume = "19",
    pages = "527--550",
    year = "2002"
}

@article{Hwang:2002fp,
    author = "Hwang, Jai-chan and Noh, Hyerim",
    title = "{Cosmological perturbations in a generalized gravity including tachyonic condensation}",
    eprint = "hep-th/0206100",
    archivePrefix = "arXiv",
    doi = "10.1103/PhysRevD.66.084009",
    journal = "Phys. Rev. D",
    volume = "66",
    pages = "084009",
    year = "2002"
}

@article{Noh:2004bc,
    author = "Noh, Hyerim and Hwang, Jai-chan",
    title = "{Second-order perturbations of the Friedmann world model}",
    doi = "10.1103/PhysRevD.69.104011",
    journal = "Phys. Rev. D",
    volume = "69",
    pages = "104011",
    year = "2004"
}

@article{Li:2024kfm,
    author = "Li, Tao and Liu, Lei-Hua",
    title = "{Inflationary Krylov complexity}",
    eprint = "2401.09307",
    archivePrefix = "arXiv",
    primaryClass = "hep-th",
    doi = "10.1007/JHEP04(2024)123",
    journal = "JHEP",
    volume = "04",
    pages = "123",
    year = "2024"
}

@article{Nojiri:2006ri,
    author = "Nojiri, Shin'ichi and Odintsov, Sergei D.",
    editor = "Borowiec, Andrzej",
    title = "{Introduction to modified gravity and gravitational alternative for dark energy}",
    eprint = "hep-th/0601213",
    archivePrefix = "arXiv",
    reportNumber = "KARP-2006-06",
    doi = "10.1142/S0219887807001928",
    journal = "eConf",
    volume = "C0602061",
    pages = "06",
    year = "2006"
}

@article{Nojiri:2017ncd,
    author = "Nojiri, S. and Odintsov, S. D. and Oikonomou, V. K.",
    title = "{Modified Gravity Theories on a Nutshell: Inflation, Bounce and Late-time Evolution}",
    eprint = "1705.11098",
    archivePrefix = "arXiv",
    primaryClass = "gr-qc",
    reportNumber = "PHYS.REPT.-692-(2017)-1-104, Phys.Rept. 692 (2017) 1-104",
    doi = "10.1016/j.physrep.2017.06.001",
    journal = "Phys. Rept.",
    volume = "692",
    pages = "1--104",
    year = "2017"
}

@article{Ali:2018fcz,
    author = "Ali, Tibra and Bhattacharyya, Arpan and Shajidul Haque, S. and Kim, Eugene H. and Moynihan, Nathan",
    title = "{Time Evolution of Complexity: A Critique of Three Methods}",
    eprint = "1810.02734",
    archivePrefix = "arXiv",
    primaryClass = "hep-th",
    reportNumber = "YITP-18-111",
    doi = "10.1007/JHEP04(2019)087",
    journal = "JHEP",
    volume = "04",
    pages = "087",
    year = "2019"
}

@article{Jefferson:2017sdb,
    author = "Jefferson, Ro and Myers, Robert C.",
    title = "{Circuit complexity in quantum field theory}",
    eprint = "1707.08570",
    archivePrefix = "arXiv",
    primaryClass = "hep-th",
    doi = "10.1007/JHEP10(2017)107",
    journal = "JHEP",
    volume = "10",
    pages = "107",
    year = "2017"
}

@article{Barbon:2019wsy,
    author = "Barb{\'o}n, J. L. F. and Rabinovici, E. and Shir, R. and Sinha, R.",
    title = "{On The Evolution Of Operator Complexity Beyond Scrambling}",
    eprint = "1907.05393",
    archivePrefix = "arXiv",
    primaryClass = "hep-th",
    reportNumber = "IFT-UAM/CSIC-19-98",
    doi = "10.1007/JHEP10(2019)264",
    journal = "JHEP",
    volume = "10",
    pages = "264",
    year = "2019"
}

@article{Chapman:2017rqy,
    author = "Chapman, Shira and Heller, Michal P. and Marrochio, Hugo and Pastawski, Fernando",
    title = "{Toward a Definition of Complexity for Quantum Field Theory States}",
    eprint = "1707.08582",
    archivePrefix = "arXiv",
    primaryClass = "hep-th",
    doi = "10.1103/PhysRevLett.120.121602",
    journal = "Phys. Rev. Lett.",
    volume = "120",
    number = "12",
    pages = "121602",
    year = "2018"
}

@article{Bhattacharyya:2018bbv,
    author = "Bhattacharyya, Arpan and Shekar, Arvind and Sinha, Aninda",
    title = "{Circuit complexity in interacting QFTs and RG flows}",
    eprint = "1808.03105",
    archivePrefix = "arXiv",
    primaryClass = "hep-th",
    reportNumber = "YITP-18-89",
    doi = "10.1007/JHEP10(2018)140",
    journal = "JHEP",
    volume = "10",
    pages = "140",
    year = "2018"
}

@article{Jiang:2018nzg,
    author = "Jiang, Jie and Liu, Xiangjing",
    title = "{Circuit Complexity for Fermionic Thermofield Double states}",
    eprint = "1812.00193",
    archivePrefix = "arXiv",
    primaryClass = "hep-th",
    doi = "10.1103/PhysRevD.99.026011",
    journal = "Phys. Rev. D",
    volume = "99",
    number = "2",
    pages = "026011",
    year = "2019"
}

@article{Alves:2018qfv,
    author = "Alves, Daniel W. F. and Camilo, Giancarlo",
    title = "{Evolution of complexity following a quantum quench in free field theory}",
    eprint = "1804.00107",
    archivePrefix = "arXiv",
    primaryClass = "hep-th",
    doi = "10.1007/JHEP06(2018)029",
    journal = "JHEP",
    volume = "06",
    pages = "029",
    year = "2018"
}

@article{Camargo:2018eof,
    author = "Camargo, Hugo A. and Caputa, Pawel and Das, Diptarka and Heller, Michal P. and Jefferson, Ro",
    title = "{Complexity as a novel probe of quantum quenches: universal scalings and purifications}",
    eprint = "1807.07075",
    archivePrefix = "arXiv",
    primaryClass = "hep-th",
    doi = "10.1103/PhysRevLett.122.081601",
    journal = "Phys. Rev. Lett.",
    volume = "122",
    number = "8",
    pages = "081601",
    year = "2019"
}

@article{Ali:2018aon,
    author = "Ali, Tibra and Bhattacharyya, Arpan and Shajidul Haque, S. and Kim, Eugene H. and Moynihan, Nathan",
    title = "{Post-Quench Evolution of Complexity and Entanglement in a Topological System}",
    eprint = "1811.05985",
    archivePrefix = "arXiv",
    primaryClass = "hep-th",
    reportNumber = "YITP-18-116",
    doi = "10.1016/j.physletb.2020.135919",
    journal = "Phys. Lett. B",
    volume = "811",
    pages = "135919",
    year = "2020"
}

@article{Khan:2018rzm,
    author = "Khan, Rifath and Krishnan, Chethan and Sharma, Sanchita",
    title = "{Circuit Complexity in Fermionic Field Theory}",
    eprint = "1801.07620",
    archivePrefix = "arXiv",
    primaryClass = "hep-th",
    doi = "10.1103/PhysRevD.98.126001",
    journal = "Phys. Rev. D",
    volume = "98",
    number = "12",
    pages = "126001",
    year = "2018"
}

@article{Zhao:2019nxk,
    author = "Zhao, Ying",
    title = "{A quantum circuit interpretation of evaporating black hole geometry}",
    eprint = "1912.00909",
    archivePrefix = "arXiv",
    primaryClass = "hep-th",
    doi = "10.1007/JHEP07(2020)139",
    journal = "JHEP",
    volume = "07",
    pages = "139",
    year = "2020"
}

@article{Brown:2019rox,
    author = "Brown, Adam R. and Gharibyan, Hrant and Penington, Geoff and Susskind, Leonard",
    title = "{The Python{\textquoteright}s Lunch: geometric obstructions to decoding Hawking radiation}",
    eprint = "1912.00228",
    archivePrefix = "arXiv",
    primaryClass = "hep-th",
    doi = "10.1007/JHEP08(2020)121",
    journal = "JHEP",
    volume = "08",
    pages = "121",
    year = "2020"
}

@article{Chandra:2021kdv,
    author = {Chandra, A. Ramesh and de Boer, Jan and Flory, Mario and Heller, Michal P. and H{\"o}rtner, Sergio and Rolph, Andrew},
    title = "{Spacetime as a quantum circuit}",
    eprint = "2101.01185",
    archivePrefix = "arXiv",
    primaryClass = "hep-th",
    doi = "10.1007/JHEP04(2021)207",
    journal = "JHEP",
    volume = "21",
    pages = "207",
    year = "2021"
}

@article{Li:2021sro,
    author = "Li, Ai-chen",
    title = "{Complexity, chaos and the moving $D_3$-brane}",
    eprint = "2110.15855",
    archivePrefix = "arXiv",
    primaryClass = "hep-th",
    month = "10",
    year = "2021"
}

@article{Zhai:2024tkz,
    author = "Zhai, Ke-Hong and Liu, Lei-Hua and Zhang, Hai-Qing",
    title = "{Generalized CV Conjecture and Krylov Complexity in Two-Mode Hermitian Systems via Information Geometry}",
    eprint = "2412.08925",
    archivePrefix = "arXiv",
    primaryClass = "hep-th",
    doi = "10.1016/j.aop.2026.170534",
    journal = "Annals Phys.",
    volume = "491",
    pages = "170534",
    year = "2026"
}

@article{Zhai:2024odw,
    author = "Zhai, Ke-Hong and Liu, Lei-Hua",
    title = "{Krylov Complexity in Early Universe}",
    eprint = "2411.18405",
    archivePrefix = "arXiv",
    primaryClass = "hep-th",
    doi = "10.1093/ptep/ptag012",
    journal = "PTEP",
    volume = "2026",
    number = "2",
    pages = "023E04",
    year = "2026"
}

@article{Liu:2025caj,
    author = "Liu, Shi-Cheng and Liu, Lei-Hua and Li, Bichu and Zhang, Hai-Qing and He, Peng-Zhang",
    title = "{A quantum information method for early universe with non-trivial sound speed}",
    eprint = "2510.04011",
    archivePrefix = "arXiv",
    primaryClass = "gr-qc",
    doi = "10.1002/prop.70081",
    journal = "Fortsch. Phys.",
    volume = "74",
    pages = "e70081",
    year = "2026"
}
\end{document}